\documentclass{article}
\usepackage{arxiv}
\usepackage[utf8]{inputenc}
\usepackage{graphicx}
\usepackage{amsmath}
\usepackage{amssymb}
\usepackage{array}
\usepackage{enumerate}
\usepackage{enumitem}
\usepackage{algorithm}
\usepackage{algpseudocode}
\usepackage{caption}

\usepackage{multirow}
\usepackage{svg}
\usepackage{times}
\usepackage{soul}
\usepackage{url}
\usepackage{float}
\usepackage{flafter}
\usepackage{latexsym}
\usepackage{color}

\DeclareMathOperator*{\argmin}{arg\,min}
\usepackage{subfigure}
\usepackage{wrapfig}

\usepackage{booktabs}
\definecolor{fixeparams}{rgb}{0,0.5,0.8}
\usepackage{hyperref}

\usepackage{xspace}
\newcommand{\xvis}{\textsc{Visual}\xspace}
\newcommand{\xtext}{\textsc{Language}\xspace}
\newcommand{\xrules}{\textsc{Rules}\xspace}
\newcommand{\xnone}{\textsc{No}\xspace}

\usepackage{tikz}
\definecolor{myorange}{rgb}{1,0.58,0}
\newcommand{\circledtext}[1]{%
    \tikz[baseline=(char.base)]{
        \node[shape=circle, fill=myorange, text=white,inner sep=1pt] (char) {#1};
    }%
}

\newcommand{\egg}{
  \tikz[baseline=-0.5ex]{
    \draw[fill=white, rotate=-20] (0,0) ellipse (0.2cm and 0.3cm); 
  }
}

\definecolor{cstronglyagree}{rgb}{0.84,0.10,0.11}
\definecolor{coloragree}{rgb}{0.99,0.68,0.38}
\definecolor{colorneitheragree}{rgb}{0.84,0.84,0.84}
\definecolor{colordisagree}{rgb}{0.67,0.85,0.91}
\definecolor{colorstronglydisagree}{rgb}{0.17,0.48,0.71}
\newcommand{\coloredsquare}[1]{%
    \tikz[baseline=0.2ex]{
        \draw[fill=#1, draw=none] (0,0) rectangle (8pt,8pt);
    }%
}

\title{Explanation format does not matter; but explanations do -
An Eggsbert \texorpdfstring{\egg}{egg} study on explaining Bayesian Optimisation tasks}

\author{
 Tanmay Chakraborty \\
  Continental Automotive Technologies GmbH, AI Lab Berlin, Germany\\
  University of Marburg, Germany\\
  \texttt{tanmay.chakraborty@continental-corporation.com} 
   \And
   Marion Koelle\\
   Hochschule RheinMain (HSRM), Wiesbaden, Germany\\
  \texttt{ marion.koelle@hs-rm.de} 
   \And
   J\"org Schl\"otterer \\
   University of Marburg, Germany \\
   University of Mannheim, Germany\\
  \texttt{joerg.schloetterer@uni-marburg.de} 
  \And
   Nadine Schlicker \\
   University of Marburg, Germany \\
  \texttt{nadine.schlicker@uni-marburg.de} 
   \And
 Christian Wirth \\
  Continental Automotive Technologies GmbH, Germany \\
  \texttt{christian.2.wirth@continental-corporation.com} 
  \And
 Christin Seifert \\
  University of Marburg, Germany \\
  \texttt{christin.seifert@uni-marburg.de} 
}

\begin{document}
\maketitle
\begin{abstract}
Bayesian Optimisation (BO) is a family of methods for finding optimal parameters when the underlying function to be optimised is unknown. BO is used, for example, for hyperparameter tuning in machine learning and as an expert support tool for tuning cyberphysical systems. 
For settings where humans are involved in the tuning task, methods have been developed to explain BO (Explainable Bayesian Optimization, XBO). 
However, there is little guidance on how to present XBO results to humans so that they can tune the system effectively and efficiently. 
In this paper, we investigate how the XBO explanation format affects users' task performance, task load, understanding and trust in XBO. We chose a task that is accessible to a wide range of users. 
Specifically, we set up an egg cooking scenario with 6 parameters that participants had to adjust to achieve a perfect soft-boiled egg. 
We compared three different explanation formats: a bar chart, a list of rules and a textual explanation in a between-subjects online study with 213 participants.
Our results show that adding any type of explanation increases task success, reduces the number of trials needed to achieve success, and improves comprehension and confidence. 
While explanations add more information for participants to process, we found no increase in user task load. 
We also found that the aforementioned results were independent of the explanation format; all formats had a similar effect.
This is an interesting finding for practical applications, as it suggests that explanations can be added to BO tuning tasks without the burden of designing or selecting specific explanation formats. 
In the future, it would be interesting to investigate scenarios of prolonged use of the explanation formats and whether they have different effects on users' mental models of the underlying system.
\end{abstract}



\section{Introduction}
\label{sec:introduction}

Bayesian Optimization (BO) is a model-based sequential optimization algorithm that is widely used for hyperparameter tuning of cyberphysical systems or in machine learning~\cite{shahriari2015taking}. 
Parameter tuning is crucial for optimizing the performance of a system or model by identifying its ideal operating parameters~\cite{feurer2019hyperparameter}. BO systematically probes a predefined parameter search space to determine the most effective parameter settings.



A human-AI collaborative approach can support the probing for promising candidates and refinement of found settings by incorporating specialized knowledge of domain experts in BO task~\cite{hvarfner2022pibo}.
However, the opaque nature of BO poses significant challenges to effective collaboration. Experts often face uncertainty regarding which parameters have been adequately optimized and which still require adjustment. This opacity reflects broader issues in AI systems, where a lack of interpretability can impede both trust and usability~\cite{rudin2022black}. Recent studies highlight that teams using black-box AI models as tools perform worse in collaborative tasks than those working with AI tools that provide interpretable explanations~\cite{senoner2024explainable}. In response, the emerging field of Explainable Bayesian Optimization (XBO)~\cite{li2019explainability} aims to demystify these optimization processes, making them more transparent and accessible for human collaborators, much like explainable artificial intelligence (XAI)~\cite{adadi2018peeking} in the general AI domain. This shift towards explainability in optimization seeks to bridge the gap between human expertise and automated systems, enhancing the effectiveness of human-AI collaboration.

While XBO methods present explanations in diverse formats — such as rule-based~\cite{chakraborty_post-hoc_nodate} or visual~\cite{rodemann2022accounting} — there has been no systematic investigation into which format is most effective. Although human-centric aspects of XAI have been extensively studied, particularly in supervised learning, recommender systems, and information retrieval, these insights cannot be directly transferred to BO tasks. BO involves sequential decision-making where the goal is to discover an unknown optimum (both for the algorithm and the human) rather than to justify a single model prediction (standard AI systems). Furthermore, while traditional AI systems typically produce single-label outputs, BO generates multi-dimensional outcomes — sets of parameter values — which users must evaluate and adjust. In real-world cyber-physical systems, these parameter sets often involve more than 25 variables, creating complex multifactorial decision-making challenges.

Despite the increasing development of XBO methods, only a small number of approaches currently exist, and none have been evaluated from a human-centric perspective. Key aspects such as usability, user understanding, and trust remain unexplored in the context of BO. Investigating these dimensions is essential to enable effective human-AI collaboration in sequential, high-dimensional optimization tasks such as those found in chemical design, materials science, and other engineering domains~\cite{WANG2022100728,neumann2019data}.

Our research addresses this gap by evaluating the utility of three explanation formats (rule-based, visual, and textual) in parameter tuning, focusing on finding the format that optimally supports human-AI collaborative tuning.  
We draw inspiration from a similar study in XAI, which examined how user expertise influences the understanding and ease-of-use of different explanation formats for a predictive AI model~\cite{10.1145/3397481.3450662}. 
However, their findings cannot be directly transferred to XBO, as explanations in the two domains differ fundamentally in nature: while XAI explains (aspects of) the reasoning of a known (in terms of functional form and parameters) AI model, XBO explains (the search space of) a surrogate model of an unknown function.
We still build on their design and adapt it to evaluate various explanation formats for parameter tuning within the XBO domain.

To have access to a wide user group, we focus on parameter tuning of an egg cooking machine with the objective to cook a soft-boiled egg. The egg cooking problem is modelled by a mathematical function with six input parameters influencing the cooking time~\cite{SHI2024104332,lersch2017towards}. Participants had to adjust the machine's parameters to obtain a perfectly soft-boiled egg. This approach allowed us to replicate the challenges of parameter tuning in a context that is easy to understand and accessible to a wide range of participants with varying levels of expertise. 

We designed a between-subjects online study with 213 participants to evaluate cooking performance across six tasks. Each task required participants to tune four to five out of six parameters
Participants were allowed five trials per task and could advance to the next task only upon successfully completing the current one or exhausting their five trials. We gamified the tasks to keep participants engaged in the tasks. After each trial, participants received feedback on their tuning, simulating real-world scenarios where experts observe system performance following adjustments. 
We set up three different experimental conditions
corresponding to three explanation formats: rule-based, visual, or textual. 
Our research focused on the following question:
\begin{enumerate}
 \item [\textbf{RQ:}] \textit{What is the effect of explanation format on user performance, perceived understandability of the recommendations, perceived understanding of the underlying system, trust in the recommendation system, and the users’ task load?}
\end{enumerate}


\section{Related Work}
Explainable Artificial Intelligence (XAI) has become an essential area of research due to the widespread use of opaque, black-box AI models in critical sectors such as healthcare, finance, justice, and engineering~\cite{adadi2018peeking,10.1613/jair.1.12228}. As these systems increasingly influence decision-making, the demand for transparency has grown, driven by user expectations~\cite{doi:10.1126/scirobotics.aay7120} and legal frameworks like the EU AI Act and GDPR~\cite{doi:10.1080/17579961.2024.2313795}. These regulations have intensified the need for transparent algorithms and technical methods to make black-box systems more interpretable~\cite{arrieta2020explainable}.

Explanations in XAI are broadly classified into ante-hoc models that are inherently interpretable and post-hoc methods that are applied after a model makes predictions~\cite{guidotti2018survey}. Prominent examples of transparent models are decision trees~\cite{Breiman1984_classification, Quinlan1986_induction} and the more recent part-prototype models~\cite{Chen2019_this_looks,Nauta2023_pipnet}, combining interpretability with representation learning.
Post-hoc methods primarily focus on feature attribution, with prominent examples such as SHAP~\cite{lundberg2019explainable}, LIME~\cite{ribeiro2016why} and a multitude of gradient attribution methods (see~\cite{Wang2024GradientBF} for a recent survey). Further examples include counterfactual explanations that identify minimal changes in the input to obtain a different prediction~\cite{Miller_2021,Nguyen2024_llms}, disentangled feature representations~\cite{Chen2016_infogan}, identification of influential training instances~\cite{Han2020_explaining} and rule-based approaches~\cite{wang2015falling, ribeiro2018anchors}.

While neither of the XAI methods is directly applicable to Bayesian Optimisation~\cite{chakraborty_post-hoc_nodate}, we draw inspiration from XAI explanations of SHAP~\cite{lundberg2019explainable} and LIME~\cite{ribeiro2016why} for the design of our explanation formats. 

\subsection{Explainable Bayesian Optimization}
\label{ssec:rel-work:xbo}
The young field of Explainable Bayesian Optimization (XBO) can be categorized into post-hoc methods and online local methods. Post-hoc methods aim to provide insights by analysing the optimization space after the optimization is complete. For instance, RX-BO~\cite{chakraborty_post-hoc_nodate} and TNTRules~\cite{chakraborty2024explainable} utilize a surrogate clustering model to generate rule- and graph-based explanations to explain the BO search space, offering actionable insights for parameter tuning. GPShap~\cite{chau2024explaining} employs feature attribution techniques to determine parameter importance within the surrogate model used in BO. Partial dependence plots (PDP)~\cite{moosbauer2024} of the BO model offer graphical explanations that illustrate the relationships between parameters, further aiding in the parameter tuning process.
In contrast, online local explainability methods, such as CoExBo~\cite{adachi2024looping} and ShapleyBO~\cite{rodemann2024explaining}, generate Shapley-based explanations for individual candidates within the optimization process. \cite{li2019explainability} constrain the search space to candidates that seem reasonable to humans, e.g., candidates in the close vicinity of already evaluated settings or interpolate between evaluated settings.

In our research, we use TNTRules, which is designed for parameter tuning, to generate our explanations.

\subsection{Evaluating Explanations}
Evaluations can be categorised as: \emph{application-grounded} (with target users in the real application), \emph{human-grounded} (with lay users on simplified proxy tasks), and functionally-grounded (with automated metrics of specific aspects)~\cite{doshi2018considerations}. Due to prohibitive cost and effort, most XAI researchers resort to functionally-grounded evaluation and few conduct user studies~\cite{10.1145/3583558}.
The same is true for XBO; of the methods discussed in section~\ref{ssec:rel-work:xbo}, only CoExBo~\cite{adachi2024looping} includes human-centred testing.
This lack of human-centred evaluation leads to two challenges: First, explanations can serve multiple purposes besides transparency, such as 
building user trust, increasing system effectiveness, persuading users to act, 
and increasing overall satisfaction~\cite{Tintarev2011}. Without human-centred evaluation, insights into most of these aspects will be missed.
Second, when designing explanations, researchers rely primarily on intuition or experience, without incorporating insights from human decision theory, psychology, philosophy, or cognitive science~\cite{adadi2018peeking,10.1145/3377325.3380623,10.1145/3290605.3300831}, whereas social science research argues that explanations should be user-centred and conversational, i.e. in natural language~\cite{MILLER20191}.

The (implicit) assumption, that explanations are interpreted correctly and that "any explanation is better than none" has been  questioned~\cite{arora2022explain,dinu2020challenging}. While visual explanations might be favoured by non-expert users~\cite{10.1145/3397481.3450662}, \cite{van2021evaluating} found slight advantages for rule-based over example-based explanations. On the other hand, saliency maps (visualizations of feature attribution) might be misinterpreted~\cite{schuff2022human} and verbalisation of saliency maps can improve upon certain aspects, but at the cost of others (e.g., increase in perceived helpfulness at the cost of faithfulness)~\cite{feldhus2023-saliency}.

In light of these mixed observations in XAI, we analyse the effect of three XBO explanation formats (\xvis, \xrules, and \xtext) and a \xnone explanation baseline on user performance, perceived understanding, trust, and task load.

\section{Explaining Bayesian Optimization}
Bayesian Optimization (BO; \cite{shahriari2015taking}) is a black-box optimization method, i.e., the relations between the parameters and the target of the optimization problem is unknown. BO learns a surrogate model for the underlying optimization problem based on already evaluated parameters and iteratively chooses and evaluates new parameters. Thus, over time the model becomes more certain and comes closer to the true optimum. 
An explanation of BO informs which parameter regions are likely to produce good results, and should or should not be further explored.
We adopt the explanations generated by TNTRules, a recent XBO method~\cite{chakraborty2024explainable}. TNTRules generates a list of rules that describes ranges of the parameters and their certainty estimates. The list is binarised into two groups ``Tune'' and ``No tune'', based on their (expected) impact on the BO result. 
More details on BO and the explanation method can be found in Appendix~\ref{appendix:sec:background}.

\begin{figure*}[thbp]
    \centering
\includegraphics[width=\textwidth,trim={3cm 3cm 0cm 0cm},clip]{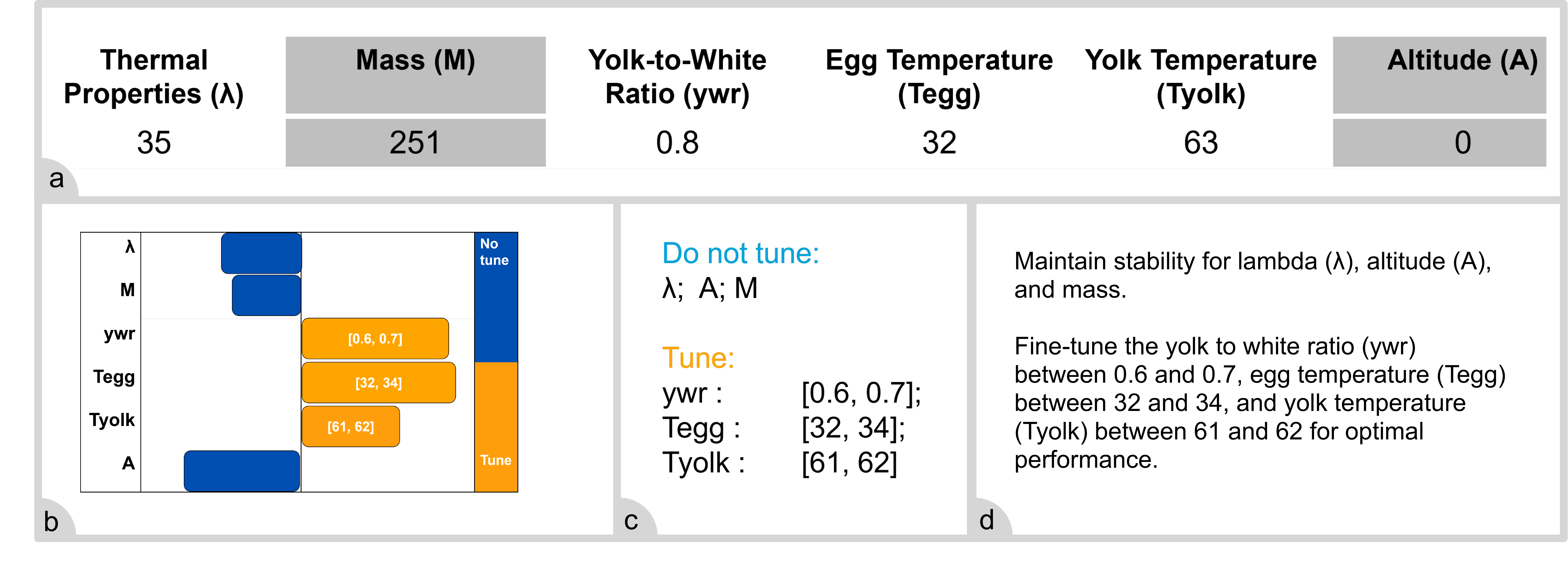}      
    \caption{Types of explanations evaluated in the user study. a) BO recommends the values shown in the table at the top. Values in cells with gray background are fixed and can not be tuned. The table is the only information shown in the ``no explanation'' condition. 
    Three different types of explanations show the same information on how to tune the parameters: b) \xvis c) \xrules, and c) \xtext.}
    \label{fig:explanationtypes}
\end{figure*}

In this study we compare three different explanation formats (see~Fig.~\ref{fig:explanationtypes}), which we describe in the following. 

\textbf{Visual}, bar-char like explanations are used by common XAI methods, such as SHAP~\cite{lundberg2019explainable} and LIME~\cite{ribeiro2016why}, and also our selected XBO method~\cite{chakraborty2024explainable}. Fig.~\ref{fig:explanationtypes} (b)) shows an example visualization. 
The visualisation lists the parameters on the y-axis) and shows their importance on the BO results on the x-axis. The order of the parameters on the x-axis is pre-defined and fixed. The binary Tune / No-Tune categorisation is shown in colors (blue indicates parameters not recommended to tune). The expected impact of the parameters on the BO results is indicated by the direction of the bar (negative x-axis direction likely worsens the outcome), and the length of the bar (to which extend the outcome is likely worsened).
Such visualisations can show continuous importance information through varying bar lengths, which is difficult to convey through e.g., rule-based explanations. 


\textbf{Rules} are discrete representations of knowledge that apply under specific conditions outlined within the rule itself~\cite{Nalepa2018}. 
The rule-based explanation consists of the \textit{antecedent}, i.e., the parameters for the optimization task and their corresponding ranges, and the \textit{consequent}, i.e., the predicted objective value with its uncertainty. Fig.~\ref{fig:explanationtypes} (c)) shows an example rule-based explanation.
The format of the rule based explanation is:
\begin{equation*} 
\begin{split}
\text{No tune}: {x_3 }, {x_4 }, 
\text{Tune}: {x_1 \in [r_{\text{lower}}^1, r_{\text{upper}}^1], x_2 \in [r_{\text{lower}}^2, r_{\text{upper}}^2]} 
\end{split}
\end{equation*}
where $x_1, x_2, ..., x_n$ are the parameters, $[r_{\text{lower}}^i, r_{\text{upper}}^i]$ represents the range of values for each parameter $x_i$, and $[y_{\text{lower}}, y_{\text{upper}}]$ is the range of the predicted objective value $y$ with $95\%$ confidence bounds, indicating the uncertainty in the BO result.
For parameters that likely do not improve the result (``no-tune'')  parameter ranges are omitted to keep the explanation concise.  

Inspired by work on verbalizing explanations (e.g.,~\cite{8933695}) we add an explanation in the form of \textbf{natural language} (cf. Fig.~\ref{fig:explanationtypes} (c)). We use GPT-4~\cite{openai2024gpt4technicalreport} to generate these explanations from the XBO output. We prompted GPT-4\footnote{We manually checked the output of GPT-4 for factual correctness.} (see prompt in Appendix~\ref{appendix:sec:textualexp}) to adhere to the following template: 
\begin{quote}
\textit{Maintain stability for $\{X_{notune}\}$.\\
Fine-tune $\{X_{tune} \in [R_{\text{lower}}^{tune}, R_{\text{upper}}^{tune}]\}$ for optimal performance.}
\end{quote}
The generated explanations contain the same information as the rule-based explanations, and used the same general categorization scheme of parameters (no-tune / tune).


\section{Apparatus}
This section describes the optimisation problem `cooking an egg' that we used for the study and the setup of the study.

\subsection{Optimization Problem and Model}
In this study, we address the optimisation problem of tuning the parameters of an egg cooking machine using a BO framework with a GP backbone~\cite{shahriari2015taking,Rasmussen2005-yj,garnett_bayesoptbook_2023}. The goal of the model is to recommend parameter settings of the egg cooker to achieve optimal cooking results. The parameters are optimised for the yolk to achieve perfect consistency, i.e. soft boiling. The BO framework iteratively selects parameter configurations that balance the exploration of new settings and the exploitation of known promising regions, effectively optimising the cooking process under uncertainty.

We chose the egg cooking scenario for two reasons. First, it can be difficult to find experts with different levels of experience in hyperparameter tuning. However, cooking is a more familiar domain to a wider audience, making it easier to find participants with relevant experience. Cooking an egg is a universally relatable activity, and tuning an egg cooker reflects the real-world challenges of parameter tuning. Secondly, cooking an egg can be defined as an optimisation problem because there is a well-defined mathematical formulation that allows us to model egg states based on the tuning of several parameters.

Egg cooking can be modelled with six input parameters, with time as output (dependent) variable. Then, the cooking time ($t_{cooked}$ in seconds) is given by $
    t_{\text{cooked}} = \lambda \cdot M^{2/3} \cdot \log\left(\frac{ywr\cdot (T_{egg} - T_{water})}{T_{yolk} - T_{water}}\right)
$
Here, the egg is treated as a uniform spherical mass ($M$) with an initial temperature ($T_{egg}$), that is immersed directly into boiling water at temperature ($T_{water}$). The parameter ($ywr$) is the yolk-to-white ratio of the egg~\cite{lersch2017towards}, and $\lambda$ denote the thermal properties of the egg. 
The egg is perfect (soft boiled) when the temperature at the yolk boundary reaches around ($T_{yolk} ~ 63^\circ C$). 
When boiling an egg anywhere in the world, 
altitude ($A$) affects the cooking process by lowering the boiling point of water. At higher altitudes, atmospheric pressure decreases, which lowers the boiling temperature and slows the transfer of heat to the egg, and eggs take longer to cook at higher altitudes. Altitude affects both the water and yolk temperatures in the above equation. The following equation describes the cooking time $t_{\text{cooked}}$ for an egg cooked at an altitude of $A$:
\begin{equation}
\label{eq:altitude}
    t_{\text{cooked}} =  \lambda \cdot M^{2/3} \cdot \log\left(\frac{ywr \times \left(T_{egg} - \frac{\left(49.161 \times \log\left(29.921 \times \left(1 - 0.0000068753 \times A\right)^{5.2559}\right) + 44.932 - 32\right) \times 5}{9}\right)}{T_{yolk} - \frac{\left(49.161 \times \log\left(29.921 \times \left(1 - 0.0000068753 \times A\right)^{5.2559}\right) + 44.932 - 32\right) \times 5}{9}}\right)
\end{equation}

\begin{wrapfigure}{r}{0.15\textwidth} 
    \includegraphics[width=0.15\textwidth]{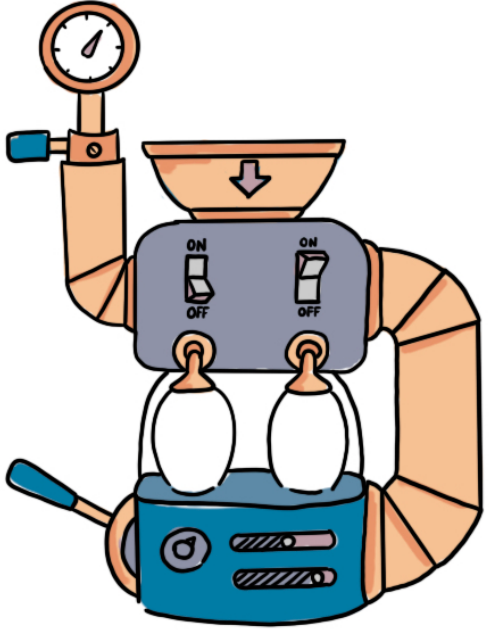}
    \caption{\\Egg cooking machine}
    \label{fig:cooking-machine}
\end{wrapfigure}
In our task, users must estimate the correct values of the input parameters mass ($M$), thermal properties ($\lambda$), yolk-to-white ratio ($ywr$), initial egg temperature ($T_{egg}$), yolk temperature ($T_{yolk}$), and altitude ($A$)
to achieve the soft-boiled state, i.e., a specific cooking time $t_{\text{cooked}}$.  
A soft boiled (chicken) egg at sea level is produced when the cooking time is between 4 min 20 sec and 4 min 45 sec.~\cite{SHI2024104332}
We extend this to allow for variations in egg size and altitude, e.g. smaller eggs need less cooking time. Note that in a real-world cooking scenario, all parameters would be fixed and a cook would have to calculate the ideal cooking time for a given egg at a given location (altitude). We have modified this scenario to virtual egg cooking, where altitude and egg properties can be changed to simulate a higher dimensional optimisation problem with non-linear dependencies. In our user study, we therefore introduce a virtual device, the egg cooking machine.

\subsection{Study Design}
We conducted a one-factor, between-subjects, fully randomised online experiment to assess the effect of explanation format on \textit{user performance}, \textit{perceived understanding}, \textit{trust} and \textit{task load}. We used explanation format as an independent variable, testing three different formats from the literature, namely \xvis, \xrules, and \xtext explanations (see Fig.~\ref{fig:explanationtypes}), along with a baseline condition of \xnone explanation. 
Each group performed the task with \xnone explanation to establish a baseline and one of the three different explanation formats (\xvis, \xrules, \xtext). 
The online experiment was administered using a custom programmed survey tool (HTML and Javascript). In order to motivate participants, we chose an engaging tone for the instructions and textual elements, and created custom illustrations (e.g., Fig.~\ref{fig:cooking-machine}). 
In accordance with local law and the policy of the senior (last) author's institution, this study did not require pre-registration with an institutional review board (IRB). It does not involve deception and is of low physical risk, i.e., no risks other than those associated with everyday life. It does not contain harmful content, address potentially triggering issues, or involve the collection of sensitive or identifiable information. In addition, all participants gave informed consent to take part.

\subsubsection{Task and Trials}

\begin{figure}
    \centering
    \includegraphics[width=\textwidth]{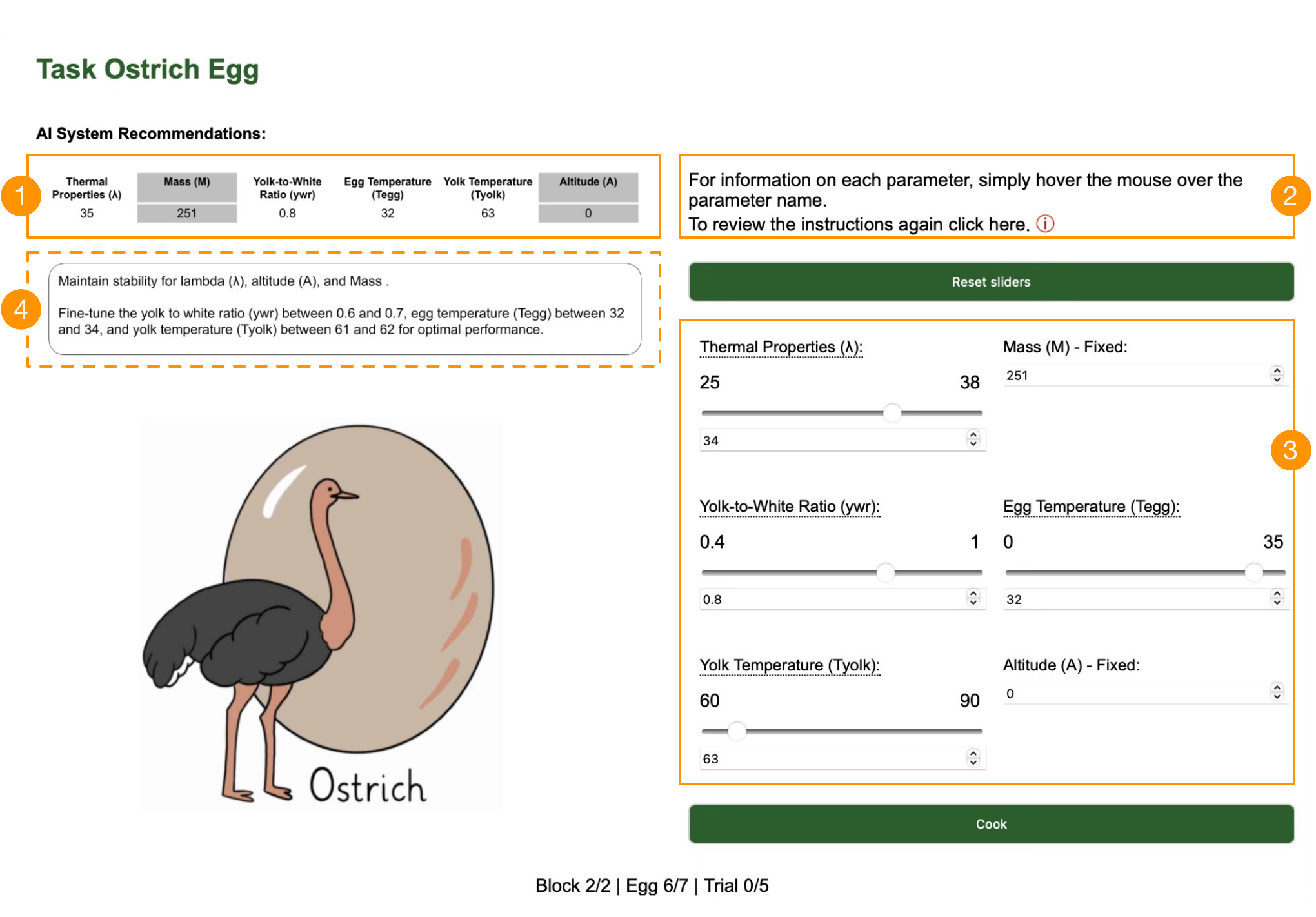}
    \caption{The user interface for the tuning consists of four main components (orange border) in the conditions with explanations, and three in the baseline condition without explanation:
1) The original recommendations of the AI in a table, showing which parameters are fixed (grey background), and which are tunable (white background). 
2) In-context-help.
3) Sliders and text fields for user input.
4) The recommendations of the advanced AI system (Only present in the conditions with explanations, here \xvis).
The sliders are set to the BO recommendations (table 1) by default. Participants need to tune the parameters to achieve a perfect, soft-boiled egg. 
 }
    \label{fig:task-overview}
\end{figure}


Participants were given the \textbf{task} of adjusting the parameters of an egg cooking machine to produce a perfect soft-boiled egg. They were given a specific type of egg and initial parameter settings, some of which were fixed and some of which were adjustable. Each task, i.e. cooking an egg, was presented on a separate page. The task interface (see Fig.~\ref{fig:task-overview}) consisted of four key elements.

The \textbf{type of egg} at the top of the page with picture of the bird and its egg set the context for the cooking scenario.
The \textbf{``AI system recommendation''} showed the output and explanations of the optimiser. The output was presented as a table with the BO-determined values for each parameter, with grey cells indicating fixed (non-tunable) parameters (Fig.~\ref{fig:task-overview} \circledtext{1}). In the \xnone explanation condition, only this table was shown.  In all three conditions with explanations, the respective explanation was added below the table (Fig.~\ref{fig:task-overview}, \circledtext{4} for \xvis).
The \textbf{in-context help} (Fig.~\ref{fig:task-overview}, \circledtext{2}) gave participants the opportunity to review the instructions and the definition of the parameters.
Sliders and text boxes allowed participants to adjust the parameters (Fig.~\ref{fig:task-overview}, \circledtext{3}). Participants could either adjust the slider or enter values into the text boxes to adjust a parameter. Sliders and text boxes were initially set to the output of the optimiser (values in the table).  We also displayed sliders and text boxes for the non-tunable parameters to provide a consistent layout across tasks, but they were not adjustable.
To ensure that the participants actively engaged in the tuning tasks, the ``Cook'' button only became active after participant made at least one adjustment to the parameters. Participants could set the parameter values back to the initial settings with the 
``Reset sliders'' button. The progress bar at the bottom of the screen showed the current block, the current egg, and how many trials  have been used for the current egg.

Each participant was given five \textbf{trials} per task. After each trial (``Cook'' button clicked) they received feedback on the egg quality indicating whether the egg was undercooked, overcooked, slightly overcooked, slightly undercooked, or perfectly cooked (cf. Appendix, Table~\ref{tab:egg-feedback}).
If participants achieved a perfect egg they proceeded to the next task without further trials. If they failed, the trial was repeated up to four times. After five unsuccessful trials the task was recorded as unsuccessful, and participants moved on to the next task.


\subsubsection{Scenarios}

\begin{table}[tb]
\centering
\caption{BO recommends parameter sets for our egg cooking tasks, tailored to specific scenarios. Acceptable ranges of each parameter (search space for tuning) is given next to each parameter. \textbf{\textcolor{fixeparams}{Bold blue}} values indicate fixed parameters; singular values suggest already optimized settings that require no further tuning, while entries with arrows represent noisy settings that need tuning, i.e., BO recommended the value before the arrow, while the correct value is shown after the arrow. Due to noise, only following BO recommendations may result in overcooked or undercooked eggs.  }
\begin{tabular}{lr cccc}  
\toprule
&&
\includegraphics[width=2.cm]{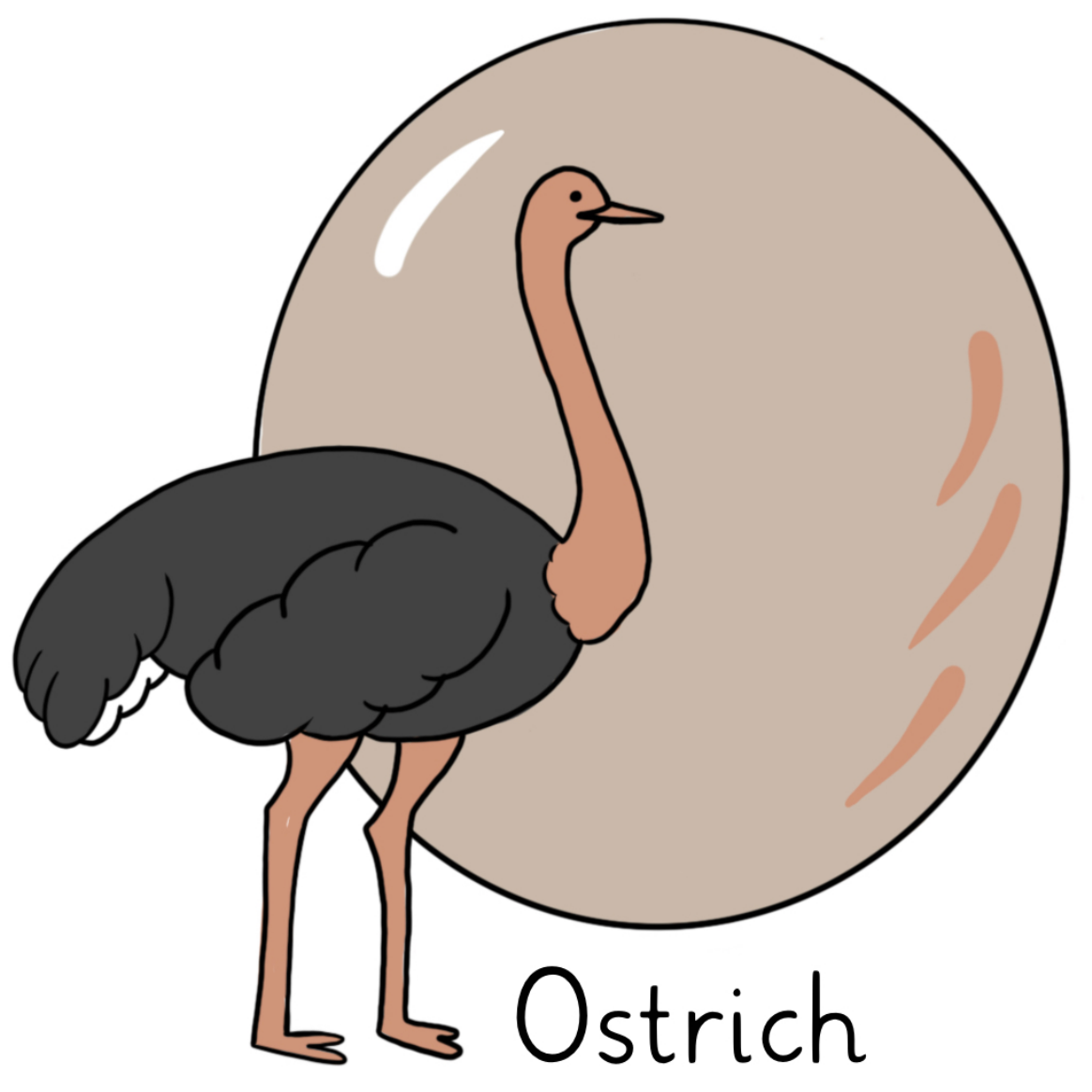}&
\includegraphics[width=2.cm]{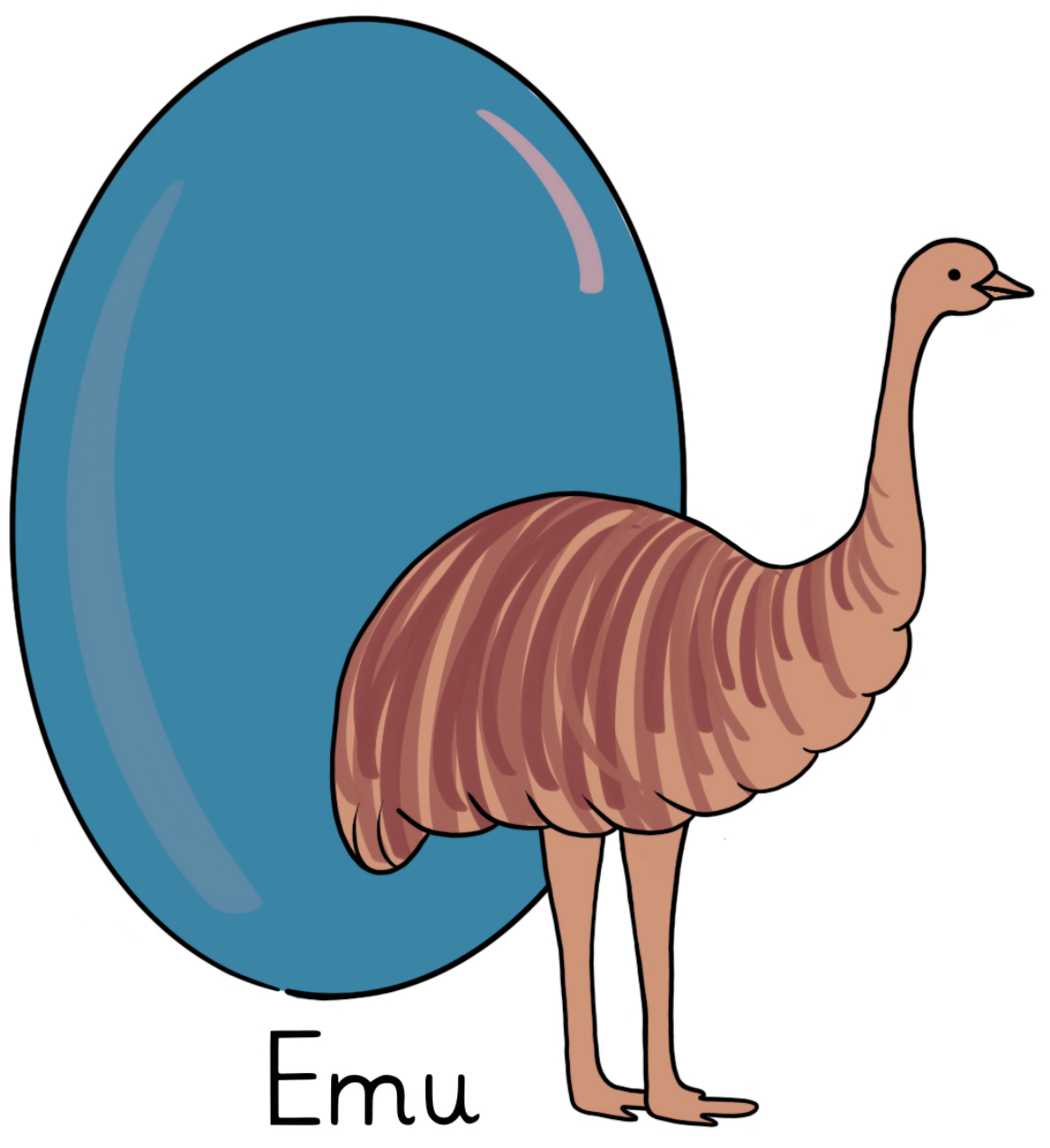}&
\includegraphics[width=2.cm]{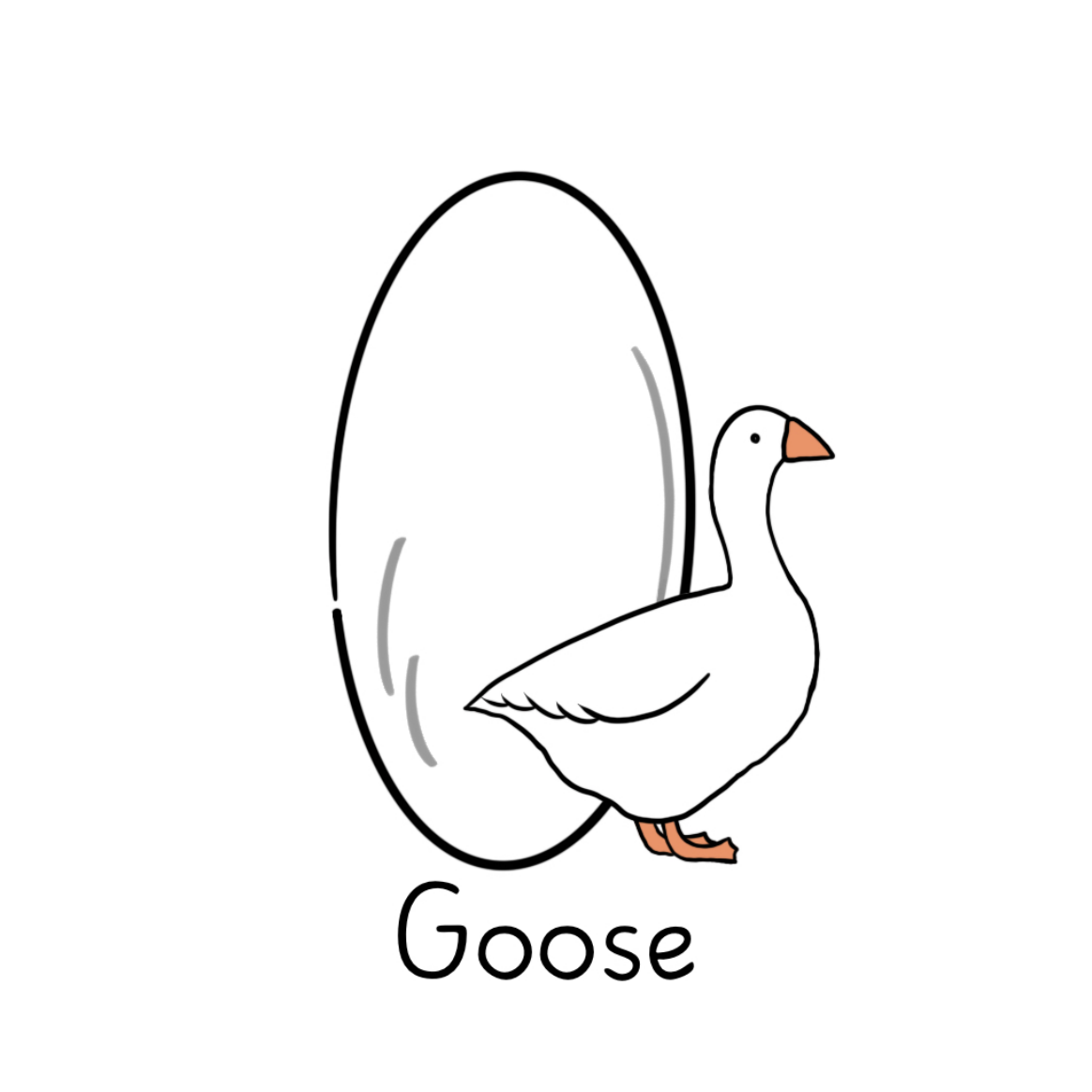}&
\includegraphics[width=2.cm]{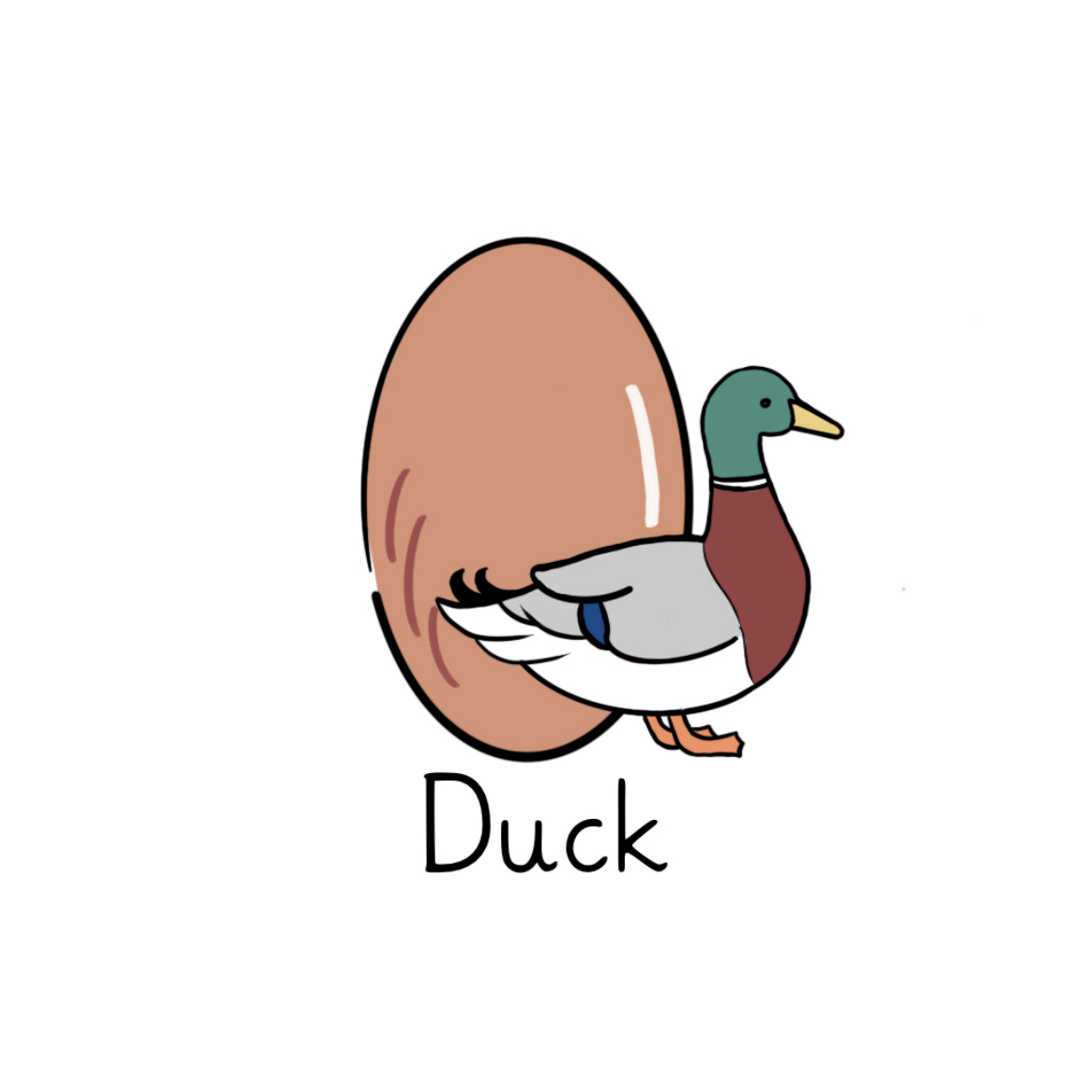}\\
\textsc{Parameters} & Egg type: & \textbf{\textcolor{fixeparams}{Ostrich}} & \textbf{\textcolor{fixeparams}{Emu}} & \textbf{\textcolor{fixeparams}{Goose}} & \textbf{\textcolor{fixeparams}{Duck}} \\
& $M$ [20, 300]: & \textbf{\textcolor{fixeparams}{251}} & $95 \longrightarrow 75$ & 75 & \textbf{\textcolor{fixeparams}{65}} \\
& $\lambda$ [25, 38]: & 35 & $27 \longrightarrow 29$ & $27 \longrightarrow 34$ & \textbf{\textcolor{fixeparams}{27}} \\
& $ywr$ [0.4, 1]: & $0.8 \longrightarrow 0.7$ & $0.6 \longrightarrow 0.9$ & \textbf{\textcolor{fixeparams}{0.5}} & $0.7 \longrightarrow 0.8$\\
& $T_{egg}$ [0, 35]:& $32 \longrightarrow 34$ & 30 & 6 & $8 \longrightarrow 13$ \\
& $T_{yolk}$ [60, 90]:& $63 \longrightarrow 62$ & \textbf{\textcolor{fixeparams}{63}} & \textbf{\textcolor{fixeparams}{63}} & 63\\
& $A$ [0, 10000]:&\textbf{\textcolor{fixeparams}{0}} & 50 & $0 \longrightarrow 10000$  & $500 \longrightarrow 0$\\
&& 
\includegraphics[width=2.cm]{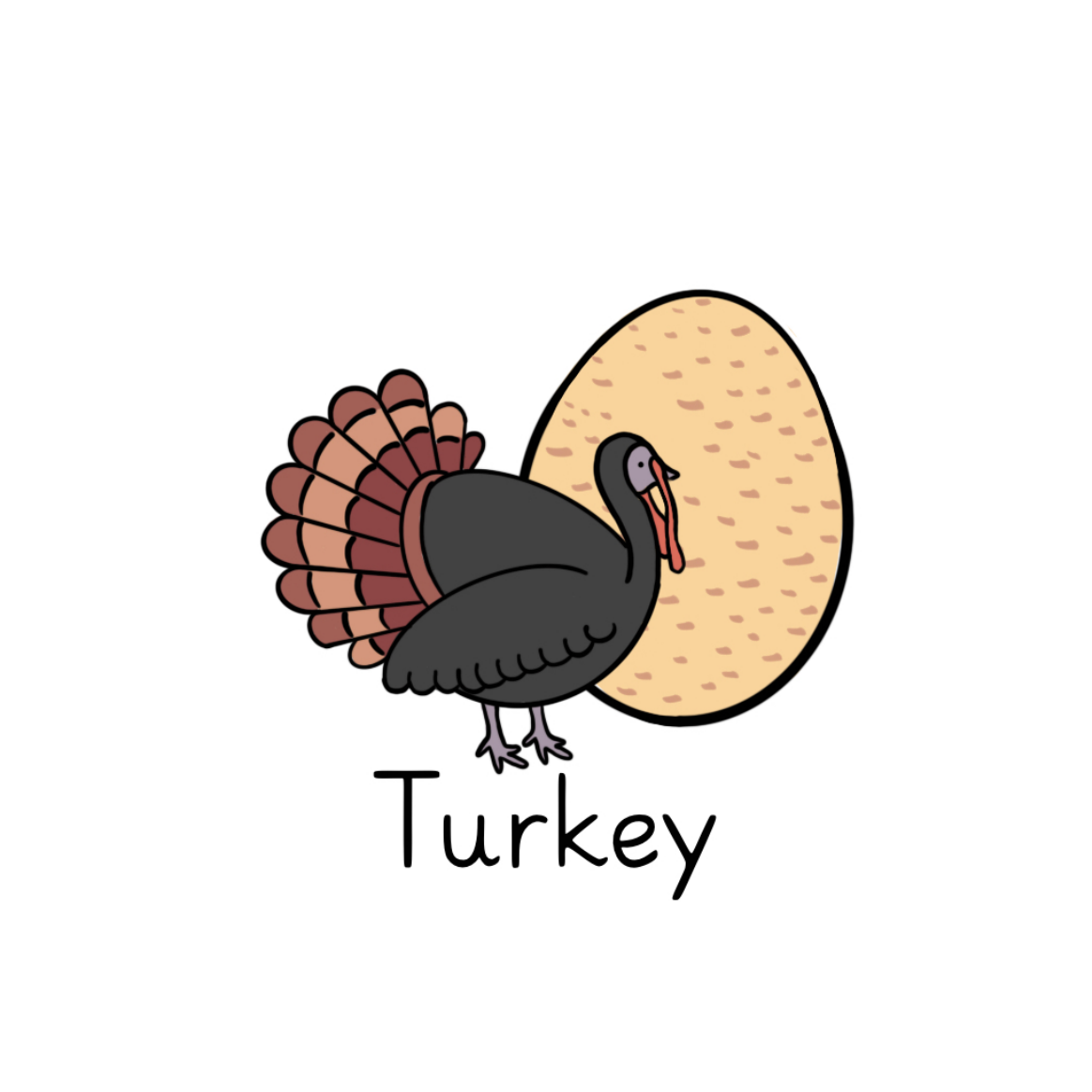}&
\includegraphics[width=2.cm]{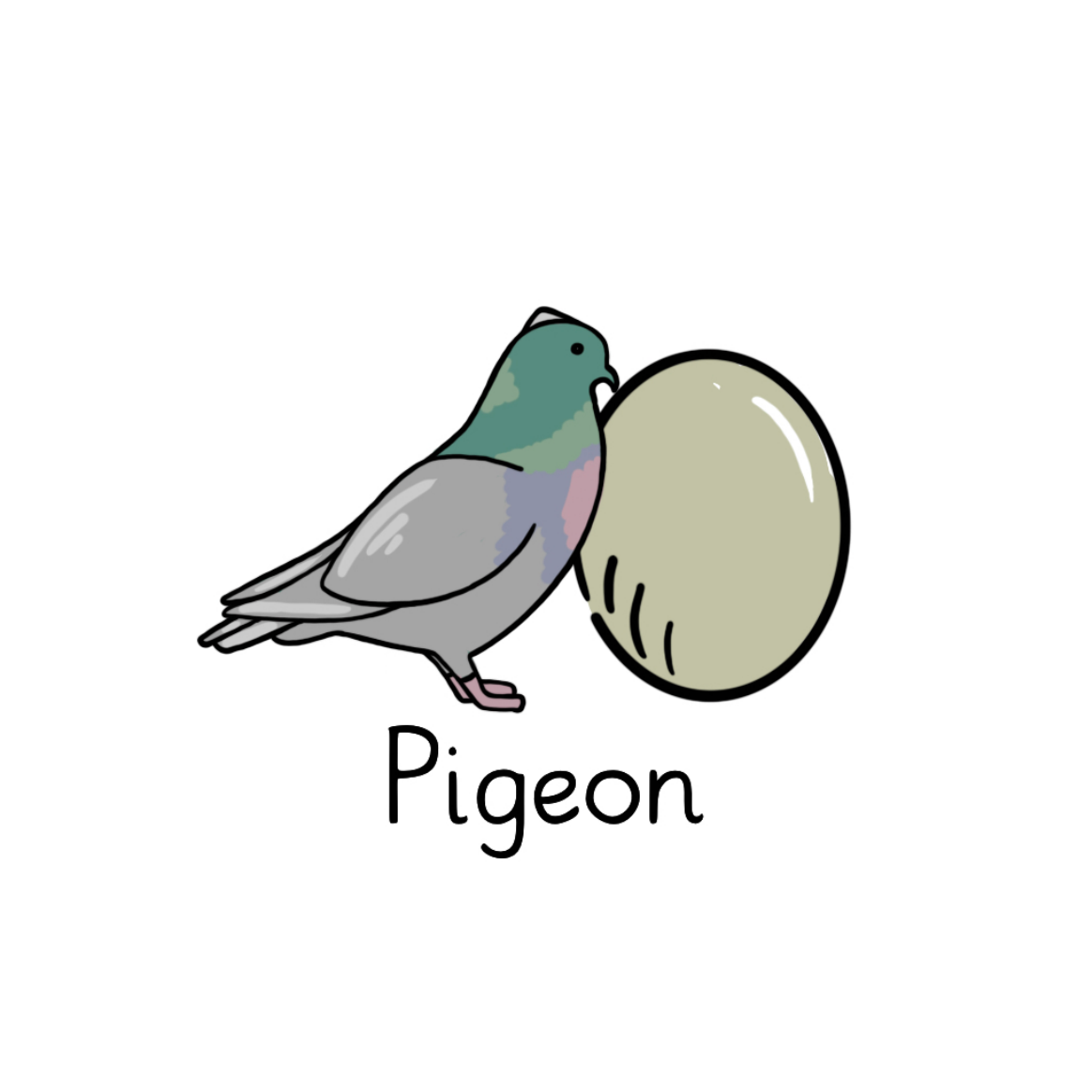}&
\multicolumn{2}{c}{\includegraphics[width=2.5cm]{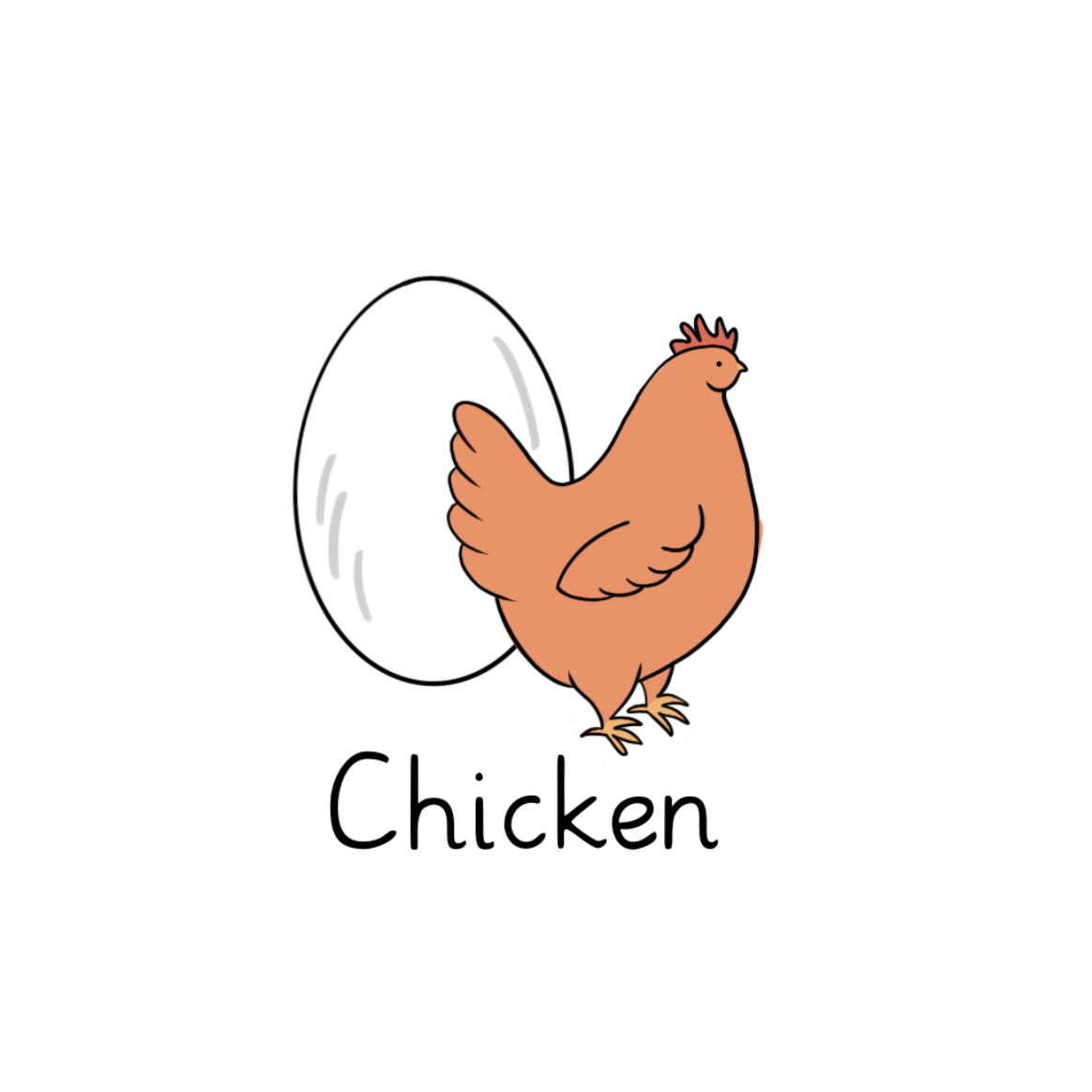}}\\
\textsc{Parameters}& Egg type: & \textbf{\textcolor{fixeparams}{Turkey}} & \textbf{\textcolor{fixeparams}{Pigeon}} & \multicolumn{2}{c}{\textbf{\textcolor{fixeparams}{Chicken}} -- Training Task}  \\
& $M$ [20, 300]:& $75 \longrightarrow 73$ & $55 \longrightarrow 62$ & \multicolumn{2}{c}{\textbf{\textcolor{fixeparams}{50}}} \\
& $\lambda$ [25, 38]:& 27 & \textbf{\textcolor{fixeparams}{27}} & \multicolumn{2}{c}{27} \\
& $ywr$ [0.4, 1]:& $0.9 \longrightarrow 0.7$ & $0.8 \longrightarrow 0.9$ & \multicolumn{2}{c}{$0.8 \longrightarrow 0.9$}\\
& $T_{egg}$ [0, 35]:& \textbf{\textcolor{fixeparams}{6}} & 21 & \multicolumn{2}{c}{12} \\
& $T_{yolk}$ [60, 90]:& 63 & \textbf{\textcolor{fixeparams}{63}} & \multicolumn{2}{c}{$60 \longrightarrow 63$}\\
& $A$ [0, 10000]:& 500 & 10 & \multicolumn{2}{c}{\textbf{\textcolor{fixeparams}{5}}}\\
\bottomrule
\end{tabular}

\label{tab:scenario_overview}
\end{table}

We created a set of seven unique scenarios, i.e., different parameter combinations with different egg types to be tuned (see Table~\ref{tab:scenario_overview}). 
Each scenario had one or two fixed parameters. Participants had to adjust the remaining parameters within an acceptable range, given as the lower and upper bounds of the parameters. 

The ranges are set following the egg cooking formula in Eqn.~\ref{eq:altitude} and general world knowledge like mass of a small egg (20g) vs. big egg (260g), temperature from the fridge ($4 ^\circ C$) vs. outside ($28 ^\circ C$), altitude near the sea side (0m) vs. mountain (below 9000m). The initial AI recommendations were set constant (as shown in the Table~\ref{tab:scenario_overview}), as they are necessary for an ecologically valid fine-tuning task.

If all parameters are adjustable, there are infinitely many ``good'' parameter settings for cooking a perfect egg. Our aim was to choose reasonable, but also diverse set of scenarios. Diversity in scenarios prevents participants from applying the same settings to multiple tasks, thus avoiding learning and memory effects. We also conducted a sensitivity analysis (details in Appendix ~\ref{appendix:parameter-sensitivity}) to identify parameters for which small changes resulted in large changes in performance. These parameters are difficult and not intuitive to tune.
For our final scenarios, we therefore i) restricted the settings to common egg variants consumed by humans, ii) ensured that each scenario had distinctly different parameter values and ranges, and iii) highly sensitive parameters rarely need to be tuned.\footnote{We keep $\lambda$ and $T_{yolk}$ stable in 5 out of 7 cases. We have added 2 scenarios with sensitive parameters, without making them the focus of most tasks, to reflect realistic conditions where sensitive parameters also need to be tuned.} The scenarios, their fixed and tunable parameters and the optimal parameter values are shown in Table~\ref{tab:scenario_overview}.
We randomly selected the scenarios and their order for participants' tasks and ensured that no participant had to cook the same egg twice.

To better replicate a realistic tuning setting, the explanations in block 2 (see Fig.~\ref{fig:task-overview} for an example) were designed such that just selecting the midpoint or simply selecting any value within the range did not represent the perfect solution. This ensured that participants were required to engage cognitively and fine-tune the system, rather than simply remaining within the given range as suggested by the explanations.

All tasks were carefully designed to prevent learning effects, and heuristics or pattern matching would not yield results due to the high variability between the tasks.

We guided participants toward achieving the ``perfect egg'' by providing feedback on the egg's state after each trial. The feedback was either ``undercooked'', ``slightly overcooked'', ``perfect'', or  ``slightly overcooked'', ``overcooked''. Appendix~\ref{appendix:egg-feedback} shows how the feedback is calculated, and Appendix, Table~\ref{tab:egg-feedback} shows the feedback images shown to participants in a pop-up window after they have submitted their parameter settings.



\subsubsection{Procedure}

After obtaining informed consent, participants were introduced to the egg cooking formula, its parameters, and the overall task of tuning an egg cooking machine (see Fig.~\ref{fig:cooking-machine}). The inclusion of the complex formula was deliberate, intended to highlight the underlying complexity of the seemingly simple task of ``egg cooking." We explicitly informed participants that the task was designed to be challenging, as extensive pilot testing had revealed the importance of setting appropriate expectations to minimize drop-out rates. To support participants during the tasks, parameter descriptions were made accessible at all times via an info icon. Additionally, during the introduction, participants were informed about the attention check embedded in the study: ``Dragon eggs are dangerous and explode when you attempt to cook them. To successfully complete this task, you should not cook a dragon egg."

Participants were then given the training task (without an explanation) of boiling a chicken egg. We chose the chicken egg as it is the most commonly used egg worldwide, assuming participants would be familiar with it.

Post trial task, we added an attention check (``Which type of egg explodes when you try to cook it and therefore must not be cooked?'', with the expected answer ``dragon egg''). 
In the first block, participants had to cook three randomly selected eggs without any explanation, with a maximum of five trials per egg. In the second block, participants had to cook another three randomly selected eggs with one of the explanation formats (\xvis, \xrules, \xtext). We also ensured that no participant was asked to cook the same egg twice. In both blocks, participants rated the task difficulty after each egg. 
After each block, participants completed the questionnaires on understanding, trust, and task load.

\subsubsection{Measures and Questionnaires}
To assess the effect of the explanation format, we used a combination of observable measures and questionnaires. 
\begin{description}
    \item[Task Performance:] We recorded the \emph{success rate} by observing whether participants achieved a perfect egg within five trials in each task (success or failure). We further measured the \emph{number of trials} participants required to achieve success (0 to maximum 4). 
    \item[Trust:] We used the Trust in Automation (TiA) questionnaire to measure trust and understanding of the AI system. TiA is particularly suited for our study as it assesses both trust and comprehension, which are crucial for evaluating user reliance on automated systems in complex tasks~\cite{korber2019theoretical}. TiA uses a 5-point Likert scale (1: strongly disagree, 2: rather disagree, 3: neither agree nor disagree, 4: rather agree, 5: strongly agree).
    \item[Understanding:]  Inspired by the structure of items in TiA, we asked two additional questions to specifically assess how understandable the AI recommendations and the cooking machine were: \emph{I was able to understand the AI recommendations for cooking the egg.}, and \emph{I was able to understand how the egg cooking machine works.} We used the same 5-point Likert scale as TiA, but analysed these questions separately as single-item scales, which look at different aspects of understanding compared to TiA subscale \textit{understanding}. Analysing them as single-item scales is appropriate in our case as the construct is unambiguous or narrow in scope~\cite{wanous1997overall}. Such items are common in XAI literature as well~\cite{10316181}.
    \item[Task Load:] We used the NASA Task Load Index (TLX) to measure participants' task load across six subscales. The TLX is widely recognized for its effectiveness in identifying task overload and stress, and therefore suitable for understanding the cognitive demands of tuning tasks~\cite{Hancock1988_nasa-tlx}. The common digital version of TLX uses a 7-point Likert scale to assess task load (1: very low, 2: low, 3: somewhat low, 4: neutral, 5: somewhat high, 6: high, 7: very high)~\cite{cezar2022domains,pedro2020development,jirapinyo2017preclinical}. 
\end{description}
Additionally, after each task, participants rated the difficulty of cooking the egg (\textbf{Task Difficulty)} using the single-item questionnaire~\cite{Sauro2009_Post-Task-Usability-Questionaire}: \textit{Overall, this task was?} on a 7-point Likert scale (1: very easy, 7: very difficult).

\subsubsection{Participants and Recruitment}

We recruited participants online via Prolific~\cite{prolific}. Prolific participants are trained to respond to surveys. They have a higher attention span, and provide more complete and honest answers~\cite{douglas2023dataquality}. Based on a power analysis assuming a small effect size $Cohen's~d \geq .2$, a standard power of .8, and an error probability of .05, we determined using Gpower~\cite{faul2007g}, a required participant pool of N=186 with 62 participants per group. Participants were pre-screened to have English C1 level. Each participant was compensated \$3.7 for approximately 25 minutes of participation according to Prolific's ethical reward policy.

\section{Results}
\label{sec:analysis}
During the study, participants cooked a total of 1278 (simulated) eggs. In the following, we detail on their background and profile, and the extend to which they succeeded in achieving a ``perfect egg'' with or without provided explanations. We further detail on subjectively perceived trust, understandability, and task load.

\subsection{Participants}
After piloting, we recruited a total of 225 participants (110F, 102M, 1D, mean age: 33; SD: 11; min age: 18; max age: 83). We excluded 12 participants (5.3\%) since they failed the attention check. This left data from 213 participants for our final analysis. 
These 213 participants were equally distributed over all three experimental conditions, namely \xvis, \xrules, or \xtext with 71 participants each. 
Details on demographics (education level, AI and explainable AI proficiency) can be found in Appendix~\ref{appendix:demographics}.
Participants indicated high proficiency in (egg) cooking: a large majority cooked \textit{very frequently}~(34\%) or \textit{frequently}~(39\%). Egg cooking is especially popular with a total of 126 participants engaging \textit{very frequently} in preparing egg-related meals. All of them had cooked eggs before (i.e., no one indicated \textit{never}). This confirms that ``cooking eggs'' is indeed a widely known, accessible and relatable scenario. Thus, it is well suited as exemplary setting to study experts completing parameter tuning tasks - here: expert egg cooks.
The tuning task was overall perceived as slightly difficult, but only the goose egg was perceived significantly harder to cook ($p<0.01$) than the trial task of cooking a chicken egg (details in Appendix~\ref{appendix:egg-difficulty}).

\subsection{Do explanations facilitate parameter tuning tasks?}

To statistically assess the effect of providing an explanation, we compare the three explanation formats (\textit{between-groups}) and data from the baseline and experimental conditions (\textit{within-groups}) in terms of (1) the total number of perfect eggs participants achieved (\textit{success rate}) and (2) the number of trials needed to cook a perfect egg (\textit{trials}). 

In the baseline condition (i.e., no explanation), participants boiled a total of 3x213=639 eggs, of which 185 were ``perfect'' (29\%), compared to a total of 344 out of 639 successfully completed tasks in the experimental condition~(54\%). We performed a Shapiro-Wilk test to ascertain that the distributions of success rate~($W=.8, p<.01$) and trials~($W=.9, p<.01$) were not normally distributed. Based on that, we employed non parametric tests.

A Wilcoxon signed-rank test indicated that the success rate was significantly higher when an explanation was given 
compared to the baseline condition with no explanation, $z=1457.0, p<.001$. On average, participants achieved a success rate of .59 (SD=.32) in the three explanation conditions compared with an average success rate of .32 (SD=.25) in the baseline condition (average success rate is the mean of individual success rates of participants). This means that they achieved 1.85 times as many perfect eggs when given an explanation. 
They also needed significantly less trials to cook a perfect egg in the experimental conditions, $z=1644.0, p<.001$:  while the three explanation types, \xvis~(M=1.56, SD=.82), \xrules~(M=1.47, SD=1.16), or \xtext~(M=1.72, SD=1.07), required a similar number of trials, significantly more trials were needed in the baseline condition (M=2.39, SD=1.36).

Yet, a Kruskal-Wallis test confirmed that there was neither a significant difference in success rate, $X^2$ (2, N=213) = 1.49, p=.47 nor trials across the three explanation formats, $X^2$ (2, N=213) = 0.22, p=.89. This indicates that all three formats, \xvis, \xrules, and \xtext, are equally effective in facilitating the egg parameter tuning task.

\subsection{Do explanations increase the user's subjective understanding?}
\begin{figure}[htbp]
    \centering
      \includegraphics[width=0.75\linewidth]{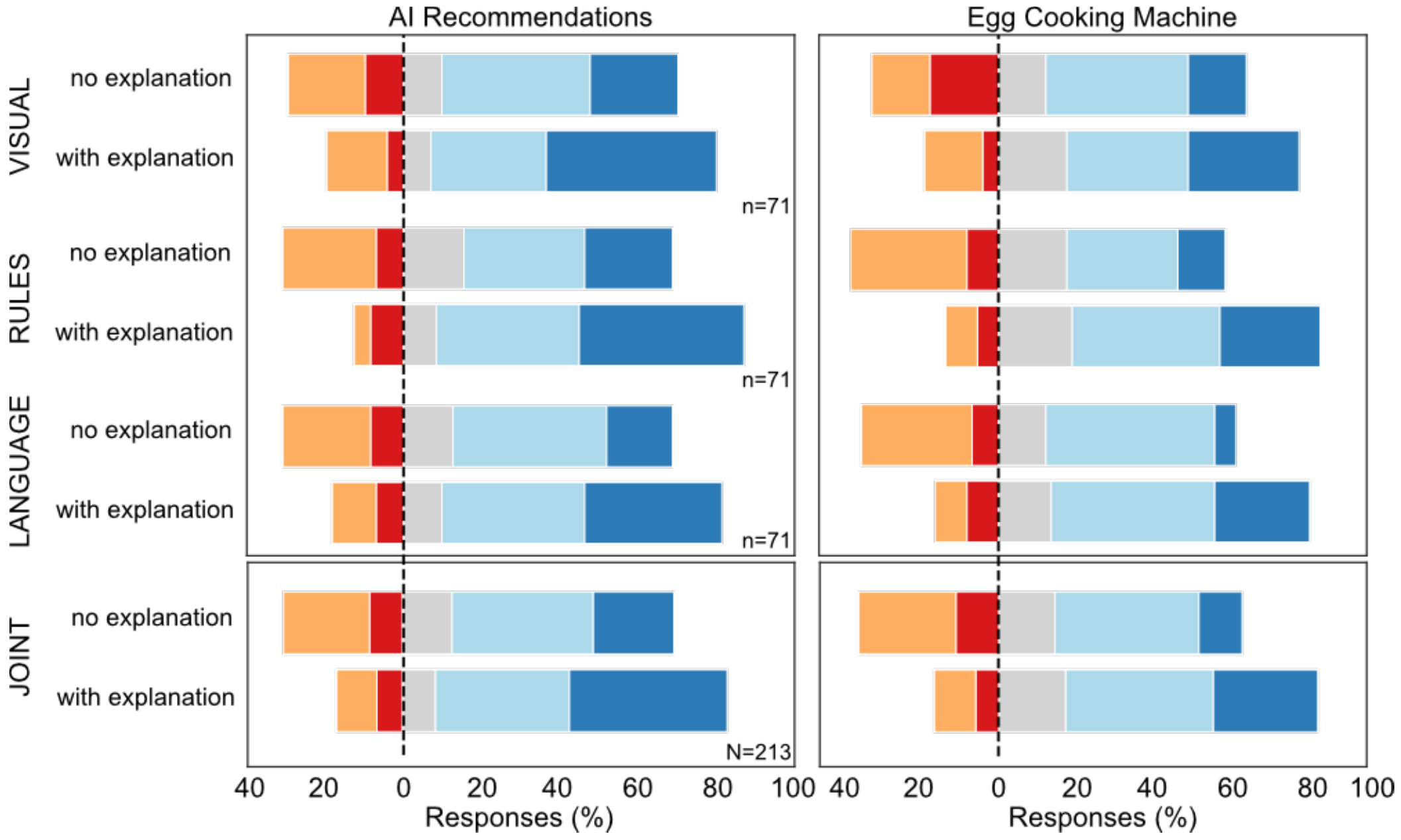}
      \caption{Explanations significantly improve perceived understanding. Chart illustrating the participants' agreement with two items structured as \textit{I was able to understand..} (1) \textit{..the AI recommendations for cooking the egg} (left) and (2) \textit{..how the egg cooking machine works} (right). Colour legend: 
\protect\coloredsquare{cstronglyagree} strongly disagree, 
\protect\coloredsquare{coloragree} rather disagree, 
\protect\coloredsquare{colorneitheragree} neither agree nor disagree, 
\protect\coloredsquare{colordisagree} rather agree,
\protect\coloredsquare{colorstronglydisagree} strongly agree.
}
\label{fig:results:understanding}
\end{figure}

We postulated further hypotheses, this time regarding the user's perceived understanding, namely whether there is a significant difference between the three explanation formats in terms of the perceived understandability of (1) the AI recommendations or (2) the underlying system. We tested normality for the two items, (1) \textit{I was able to understand the AI recommendations for cooking the egg}~($W=.8, p<.01$) and
(2) \textit{I was able to understand how the egg cooking machine works}~($W=.8, p<.01$) using the Shapiro-Wilk test. We found the data to significantly depart from a normal distribution and followed-up with the non-parametric Wilcoxon-signed rank test. This choice is adequate because Likert scales (here as 5-pt format from 1: strongly disagree to 5: strongly agree) inherently contain ordered categories that represent levels of agreement. Since distances between categories are unequal, they yield ordinal, but not interval data~\cite{Kaptein2012_statistical-analysis-methody-for-CHI, Robertson2012_likert-type-scales}. 

Our results show that including any explanation significantly improved the user's perceived understanding of the recommendation~($z=2498.0, p<.001$) and the underlying system~($z=1683.0, p<.001$). While the mean ratings in the baseline condition were close to neutral (\textit{neither agree nor disagree}) for both, the participants' understanding of (1) the AI recommendation (M=3.38 ,SD=1.27), and (2) the underlying system (M=3.12, SD=1.23), they shifted closer to \textit{agree} in the experimental conductions: M=3.92, SD=1.22 for item (1) and M=3.70, SD=1.16 for item 2 (cf. Fig~\ref{fig:results:understanding}).

Again, Kruskal-Wallis tests showed no significant differences between the three formats, i.e., \xvis~(M=3.9, SD=1.2), \xrules~(M=4, SD=1.2) and \xtext~(M=3.82, SD=1.2) contributed equally to participants' understanding of the recommendation~($X^2$ (2, N=213) = 1.10, p=.58). \xvis~(M=3.7, SD=1.2), \xrules~(M=3.7, SD=1.1) and \xtext~(M=3.7, SD=1.2) also seems to equally supported their understanding of the underlying system~($X^2$ (2, N=213) = 1.15, p=.93). 

\subsection{Do explanations increase user's trust in the system?}

\begin{table}[th]
  \caption{TiA subscales results comparing explanation types means and SDs. \textbf{Bold} shows statistically significant differences ($p < .05$). Results show a significant increase in trust (Total Sum Score) for all explanation formats. \xrules showed significant results in five out of six subscales, followed by \xvis three out of six, and \xtext one out of six.}
  \label{tab:tia_full}
  \centering
\resizebox{\textwidth}{!}{
  \begin{tabular}{lllllllll}
    \toprule
    \textbf{Explanation} & \textbf{Phase} & \textbf{R/C} & \textbf{U/P} & \textbf{FAM} & \textbf{IoD} & \textbf{PoT} & \textbf{TiA} & \textbf{Total Sum Score} \\ 
    \midrule
    \xrules & Baseline & \textbf{M=3.2, SD=0.5}  & M=3.2, SD=0.5  & \textbf{M=2.0, SD=0.9}  & \textbf{M=3.3, SD=0.8}  & \textbf{M=2.7, SD=0.6}  & \textbf{M=2.8, SD=1.0}  & \textbf{M=56.4, SD=8.0}  \\
    \xrules & Treatment & \textbf{M=3.4, SD=0.6}  & M=3.2, SD=0.5  & \textbf{M=2.4, SD=1.1}  & \textbf{M=3.3, SD=0.8}  & \textbf{M=2.9, SD=0.6}  & \textbf{M=3.2, SD=1.0}  & \textbf{M=59.9, SD=9.7}  \\
    \midrule
    \xvis & Baseline & M=3.4, SD=0.6  & \textbf{M=3.3, SD=0.7}  & \textbf{M=2.2, SD=1.1}  & M=3.5, SD=0.8  & M=3.0, SD=0.6  & \textbf{M=3.3, SD=1.0}  & \textbf{M=60.2, SD=10.7}  \\
    \xvis & Treatment & M=3.5, SD=0.6  & \textbf{M=3.4, SD=0.6}  & \textbf{M=2.5, SD=1.2}  & M=3.5, SD=0.8  & M=3.0, SD=0.7  & \textbf{M=3.5, SD=1.0}  & \textbf{M=62.8, SD=10.5}  \\
    \midrule
    \xtext & Baseline & M=3.4, SD=0.6  & M=3.3, SD=0.6  & \textbf{M=2.1, SD=1.1}  & M=3.4, SD=0.8  & M=2.8, SD=0.7  & M=3.2, SD=1.0  & \textbf{M=59.3, SD=9.7}  \\
    \xtext & Treatment & M=3.5, SD=0.6  & M=3.3, SD=0.5  & \textbf{M=2.2, SD=1.2}  & M=3.4, SD=0.9  & M=2.9, SD=0.7  & M=3.4, SD=1.1  & \textbf{M=60.9, SD=11.0}  \\
    \midrule
    Combined & Baseline & \textbf{M=3.3, SD=0.6}  & M=3.3, SD=0.6  & \textbf{M=2.1, SD=1.0}  & M=3.4, SD=0.8  & \textbf{M=2.8, SD=0.6}  & \textbf{M=3.1, SD=1.0}  & \textbf{M=58.6, SD=9.6}  \\
    Combined & Treatment & \textbf{M=3.4, SD=0.6}  & M=3.3, SD=0.6  & \textbf{M=2.4, SD=1.2}  & M=3.4, SD=0.8  & \textbf{M=3.0, SD=0.7}  & \textbf{M=3.4, SD=1.5}  & \textbf{M=61.2, SD=10.4}  \\ 
    \bottomrule
  \end{tabular}}
\end{table}

We used the TiA questionnaire to assess trust. TiA is scored on a 5-point Likert scale from 1: strongly disagree to 5: strongly agree. With the inverted items 5, 7, 10, 15, and 16, higher numerical scores indicate higher trust~\cite{korber2019theoretical}. We confirmed non-normality~($W= .9, p<.01$) and followed-up with an analysis of the whole (sum scored) TiA and the individual subscales. The descriptive statistics are in Table~\ref{tab:tia_full}.

Across the three groups (\xvis, \xrules, or \xtext), the baseline condition (no explanation) consistently scored lower than the experimental condition (with explanation) on all subscales except for \textit{Intention of Developers} and \textit{Understanding/Predictability}. Our data indicates that adding an explanation can significantly increase perceived \textit{Reliability/Competence}~($z=5545.0, p<.001$), \textit{Familiarity}~($z=933.0, p<.001$), \textit{Propensity to Trust}~($z=3658.0, p<.001$), and \textit{Trust in Automation}~($z=3259.5, p<.001$). We also found a significant increase in the overall TiA sum score~($z=5332.5, p<.001$), indicating that systems that provide an explanation are more trusted overall.

A Kruskal-Wallis tests did not indicate significant differences across the three explanation types~($X^2$ (2, N=213) = 2.7, p=.26 regarding the TiA score. This also holds true when testing the subscales individually. Detailed Likert scale plotted results can be found in Appendix~\ref{appendix:trust}.

\subsection{Do explanations affect the user's task load?}
\begin{table}[th]
  \caption{NASA-TLX subscales results comparing explanation types means and SDs. \textbf{Bold} indicates significant differences ($p < .05$). Significant reduction in mental demand and performance improvement for all explanation formats. \xrules show a reduction in frustration, and \xrules and \xvis show a reduction in effort as well.}
  \label{tab:nasa_tlx_reduced}
  \centering
  \scalebox{0.8}{
  \begin{tabular}{llllll}
    \toprule
    \textbf{Explanation} & \textbf{Phase} & \textbf{Mental Demand} & \textbf{Performance} & \textbf{Effort} & \textbf{Frustration} \\ 
    \midrule
    \xrules & Baseline & \textbf{M=64.1, SD=24.5}  & \textbf{M=33.8, SD=24.4}  & \textbf{M=60.8, SD=22.9}  & \textbf{M=49.5, SD=31.6}  \\
    \xrules & Treatment & \textbf{M=59.4, SD=26.2}  & \textbf{M=53.1, SD=26.2}  & \textbf{M=54.5, SD=25.5}  & \textbf{M=40.4, SD=31.7}  \\
    \midrule
    \xvis & Baseline & \textbf{M=66.0, SD=27.7}  & \textbf{M=37.8, SD=27.9}  & \textbf{M=66.7, SD=26.4}  & M=47.7, SD=33.5  \\
    \xvis & Treatment & \textbf{M=59.6, SD=30.3}  & \textbf{M=56.6, SD=28.1}  & \textbf{M=57.0, SD=24.5}  & M=39.9, SD=36.2  \\
    \midrule
    \xtext & Baseline & \textbf{M=64.3, SD=21.7}  & \textbf{M=40.6, SD=27.0}  & M=65.5, SD=19.4  & M=50.7, SD=29.6  \\
    \xtext & Treatment & \textbf{M=56.8, SD=28.1}  & \textbf{M=53.5, SD=30.7}  & M=61.5, SD=22.6  & M=44.6, SD=31.6  \\
    \midrule
    Combined & Baseline & \textbf{M=64.8, SD=24.6}  & \textbf{M=37.4, SD=26.5}  & \textbf{M=64.3, SD=23.1}  & M=49.3, SD=31.5  \\
    Combined & Treatment & \textbf{M=58.6, SD=28.2}  & \textbf{M=54.4, SD=28.3}  & \textbf{M=57.7, SD=24.3}  & M=41.6, SD=33.2  \\ 
    \bottomrule
  \end{tabular}}
\end{table}

We investigated whether the explanation format affects the user’s task load, as measured by (1) the six subscales of the NASA-TLX (a. mental demand, b. physical demand, c. temporal demand, d. performance, e. effort, and f. frustration) and (2) the combined raw TLX index (a single task load score ranging from 0 to 100)~\cite{hart2006nasa}.

We confirmed that both, the six NASA-TLX subscales ($W=.9, p<.01$) and the raw TLX index ($W=.9, p<.01$) were not normally distributed. We provide descriptive statistics in Table~\ref{tab:nasa_tlx_reduced}, the full table can be found Appendix~\ref{appendix:task-load}. 


Additionally, we conducted Wilcoxon-signed rank tests which indicate that adding explanations significantly reduce mental demand ($z=1459.5, p<.001$) and effort ($z=2975.5, p<.001$). Participants experienced their performance as significantly better using any of the explanation formats ($z = 2060, p<0.001$). Additionally, the felt less frustrated when using \xrules ($z=257.5, p=.005$), and felt less effort to be required using \xrules ($z=403.0, p=.036$) and \xvis ($z=177.0, p<0.001$). Unsurprisingly, there were no significant difference for physical demand, but also none for temporal demand, and the raw TLX index (see Appendix~\ref{appendix:task-load} for all conditions). 

A Kruskal-Wallis test did not indicate significant differences across the three explanation types for the raw TLX index ($X^2$ (2, N = 213) = 0.21, p=.89) or any of the subscales. 

These results suggest that the presence of explanations, regardless of the explanation format, does not significantly increase the overall task load. On the contrary, we found significant improvements on some of the subscales, most notably a reduction in mental load.

\section{Discussion}
\label{sec:discussion}

The study demonstrates that providing explanations in BO significantly enhances task performance, user understanding, and trust, without additional cognitive burden. We observed this consistently across the three explanation formats tested — \xvis, \xrules, and \xtext — and no significant differences were found between formats.

It is particularly noteworthy that explanations improved outcomes without increasing task load. BO often operates in complex, cognitively demanding settings where users are already required to process substantial amounts of information. Our results suggest that well-designed explanations can be integrated into such BO workflows without overwhelming the user, addressing a common concern in the application of explainable AI to real-world systems~\cite{liao2022humancenteredexplainableaixai,kim2024human}.

Moreover, the absence of significant differences between explanation formats indicates that BO explanations are robust to variations in their presentation. This opens the possibility for flexible, user-centered explanation design: developers can prioritize user preferences, context, or deployment constraints when choosing how to present explanations, without compromising effectiveness. Personalized explanation strategies, where users can select the format that best fits their cognitive style or task needs, could therefore be a viable and valuable design approach~\cite{nimmo2024user}. 

Finally, our findings underline the accessibility and utility of BO explanations. By improving understanding and trust, explanations help bridge the gap between users and the underlying optimization process, enabling more informed and confident decision-making. In safety-critical or high-stakes applications, where trust in AI recommendations is crucial, such support could significantly enhance the usability and acceptance of BO-based systems~\cite{adachi2024looping}.

XAI literature has highlighted challenges in achieving appropriate user reliance on AI systems. Overreliance, where users accept AI recommendations without sufficient scrutiny, can lead to errors of commission, especially when the AI is incorrect. Conversely, underreliance, where users disregard accurate AI advice, results in errors of omission and can be even more detrimental to performance~\footnote{\url{https://www.microsoft.com/en-us/research/wp-content/uploads/2022/06/Aether-Overreliance-on-AI-Review-Final-6.21.22.pdf}}. Studies have shown that explanations alone do not always mitigate these issues. The effectiveness of explanations depends on factors like task difficulty and the cognitive effort required to process the explanation~\cite{vasconcelos2023explanationsreduceoverrelianceai}. Similarly, example-based explanations might increase overreliance compared to feature-based ones~\cite{casolin2024evaluatinginfluencesexplanationstyle}. These findings suggest that the design and context of explanations are crucial for user reliance.

In our study, we observed that for all successful results 55.7\% of participants fully adhered to the provided explanations (only tried values within the given range), while 44.3\% deviated and explored alternative values. We found no evidence of over-reliance in explanations in our setting. 

Previous work also raised concerns about whether any explanation is better than none in AI systems~\cite{arora2022explain,dinu2020challenging}. Our results for XBO demonstrate the usefulness of explanations in improving both system performance and user perception. Explanations addressed the often-cited trust gap in XAI~\cite{papenmeier2022complicated,rudin2019stop,ribeiro2016why}. These results are consistent with the findings of ~\cite{senoner2024explainable}, highlighting the general effectiveness of explanations in improving performance in human-AI collaboration.

The assumptions made in previous work~\cite{rodemann2024explaining,adachi2024looping,chakraborty_post-hoc_nodate,chakraborty2024explainable,grushetskaya2024hpexplorer} about the need for explanations in BO are confirmed by our study. Our results show that explanations not only helped users understand the system's reasoning, but also fostered trust by increasing transparency. However, it is important to acknowledge that the increased trust observed among participants may have been influenced by the high fidelity of the explanations, i.e., the explanations were true to the system they were intended to explain.

To further substantiate these findings, it would be interesting to include more objective measures of understanding beyond self-reported perceptions.

In our study, each participant was exposed to only one explanation format, so we did not account for or measure user preferences. However, not assigning a preferred explanation format did not negatively affect performance; participants performed well regardless of the explanation format they received. This contrasts with previous research in the AI domain, where users often preferred specific explanation formats, such as visual~\cite{10.1145/3397481.3450662} or textual~\cite{rago2024exploring}. Our results suggest that explanation effectiveness is task dependent, and in the context of BO for parameter tuning, explanation format does not play a decisive role in the results. 


\subsection{Limitations and Practical Implications}
While our study provides strong evidence for the usefulness of explanations in BO, several practical limitations warrant further research.  First, the use of the egg cooking task as a proxy for parameter tuning may limit the generalisability of our findings to more complex industrial tuning tasks. Future studies should explore higher-dimensional, real-world scenarios to validate our findings. Second, the reliance on self-reported measures of understanding and trust has some limitations. Participants' perceived understanding and confidence may have been influenced by the positive results they achieved, rather than a true understanding of the system. More rigorous, objective measures of understanding would be valuable in distinguishing between real and perceived benefits of explanations. Third, our design only measured performance and perceptions after three repetitions of the tuning task. It is unclear whether one explanation format is better suited to learning over time. For example, repeated exposure to explanations may allow users to develop a clear mental model of the system, eventually reducing their reliance on explanations. This might reveal differences between formats that were not apparent in our study. Finally, we did not measure task completion times, which, while less critical in cyber-physical systems where execution time is often the primary bottleneck, may provide additional insight into the effectiveness of different explanation formats.

\section{Conclusion}
\label{sec:conclusion}
Driven by the need to understand beneficial human-AI collaboration for tuning cyberphysical systems, we designed an artificial scenario to understand the impact of explanations on tuning outcomes. 
We identified a tuning task that most humans would be familiar with, and created an artificial egg cooking machine with 6 tunable parameters based on a realistic mathematical model for cooking a `perfect' egg. 
Our egg cooking machine was equipped with explanations in different formats:  \xvis, \xrules and \xtext, which we derived from common explanation formats in explainable AI. 

We ran a between-subjects study with 213 valid responses on the Prolific platform, which -- anecdotally -- participants found very engaging and interesting (
\emph{[..]I thought a couple of more chances and I could have nailed the perfect egg! (P6)}, \emph{Though it was a fun task, I wish I had done better.} (P33))

Our results show that when explaining BO tasks with truthful explanations, the explanation format does not matter: we found no difference in effect on user confidence, task load, task performance and comprehension.
However, we did find a significant effect of explaining vs. not explaining in BO tasks, confirming a general agreement in the XAI domain. However, our explanations were designed to be truthful, i.e. to approximate the underlying system very well, and participants were unlikely to have had prior expectations about which parameters should be important and which should not, reducing the risk of expectation mismatch. 

This is an interesting finding for practical applications, as it suggests that explanations should be added to BO tuning tasks; however, there is no need to design or select specific explanation formats to achieve the best tuning results. Rather, the practitioner could follow user preferences and optimise for user satisfaction. 

In the future, it would be interesting to investigate scenarios of prolonged use of the explanation formats and to adapt this study to real use cases with domain experts.

\section*{Acknowledgments}

This study was supported by BMBF Project hKI-Chemie: humancentric AI for the chemical industry, FKZ 01|S21023D, FKZ 01|S21023G and Continental AG.

\section*{Appendix}
\begin{appendix}
\section{Background}
\label{appendix:sec:background}
In this section, we describe the Bayesian Optimization (BO) and TNTRules algorithm in detail.

\subsection{Bayesian Optimization (BO)}
BO is a sequential optimization method that relies on a probabilistic model, commonly a Gaussian Process (GP). It leverages the exploration/exploitation trade-off, incorporating uncertainty to minimize the required evaluations to find an optimal~\cite{garnett_bayesoptbook_2023}.

BO aims to locate the minimum ($\mathbf{x_{opt}}$) of a black-box objective function $f(\mathbf{x})$ within a defined and bounded search space $\mathcal{D} \subseteq \mathcal{R}^d$: $\mathbf{x_{opt}} = \argmin_{\mathbf{x} \in \mathcal{D}} f(\mathbf{x})$. Here, $f$ is an expensive-to-evaluate black-box function that is noisy and not expressed in closed form. The GP acts as a surrogate model to approximate this function~\cite{shahriari2015taking}. 

A GP is a collection of random variables where any finite subset of these variables follows a multivariate normal distribution. The GP describes a distribution over functions, characterized by a mean function $m(\mathbf{x})$ and a covariance function, also known as the kernel $k(\mathbf{x}, \mathbf{x'})$. These two components define the properties of the functions $f(\mathbf{x})$ at any location $\mathbf{x}$. The mean function is expressed as $m(\mathbf{x}) = \mathbb{E}[f(\mathbf{x})]$, while the kernel function is defined as $k(\mathbf{x}, \mathbf{x'}) = \mathbb{E}[(f(\mathbf{x}) - m(\mathbf{x}))(f(\mathbf{x'}) - m(\mathbf{x'}))]$. Together, they specify the GP as $f(\mathbf{x}) \sim GP(m(\mathbf{x}), k(\mathbf{x}, \mathbf{x'}))$~\cite{Rasmussen2005-yj}.

BO fundamentally differs from standard AI systems, which typically rely on a predefined dataset for training and evaluation. Standard XAI methods operate under the assumption that explanation frameworks have access to a static dataset, often comprising training data, ground-truth labels, and model predictions. In contrast, BO does not start with such a dataset. Instead, it dynamically generates data by sampling points from the search space $\mathcal{D}$ during the optimization process~\cite{shahriari2015taking}. This dynamic nature of BO introduces several challenges: the data collected is limited in quantity, heavily biased towards regions near the optimum, and inherently unsuitable for conventional explanation methods. These differences make BO a unique case, requiring the development of specialized XAI algorithms tailored to its dynamic and data-scarce nature.

\subsection{TNTRules Algorithm}
TNTRules is an algorithm designed to extract interpretable insights from BO processes Alg.\ref{tntrules}~\cite{chakraborty2024explainable}. It identifies key patterns in the BO solution space, clusters data, and generates human-readable rules that explain the optimization results. The algorithm comprises the following main steps:

\begin{algorithm}[htb]

\caption{TNTRules Algorithm}\label{tntrules}
\begin{algorithmic}[1]
\Require {BO search space:~$\mathcal{D},$ \\
No. of explanation samples:~$N_e,$ \\
Clustering threshold:~$t_s$} \\
Interestingness threshold:~$t_\alpha$
\Procedure{$TNTRules$}{$\mathcal{D} ,N_{e}, t_s$}
\State \textbf{\# Explanation dataset generation}
\State $\mathbf{X_{e}} \gets \{x_1, \ldots, x_{N_e}\} \sim \mathcal{U}(\mathcal{D})$ 
\State $\boldsymbol{\mu}, \boldsymbol{\sigma_y} \gets GP_{predict}(\mathbf{X_{e}})$
\State $\mathbf{E} \gets [\mathbf{X_{e}} ;\boldsymbol{\mu} ; \boldsymbol{\sigma_y}]$ 
\State \textbf{\#Clustering, Variance pruning, and Rule construction}
\State $\mathbf{E_{link}} \gets Clustering(\mathbf{E})$
\State $\mathbf{K} \gets VariancePruning(\mathbf{E_{link}}, t_s)$
\State $\rho_{\dashv}, \rho_{\vdash} \gets RuleConstruction(\mathbf{E}, \mathbf{K})$ 
\State \textbf{\# Rule ranking and filtering}
\For {$i \  in\  \rho_{\dashv}$}
\State $\rho_{temp}^i \gets find(\mathbf{X_{e}} \in [\rho_{\dashv}^i ; \rho_{\vdash}^i])$ 
\State $Rel^i \gets max(likelihood(GP_{predict}(\rho_{temp}^i)))$ 
\State $Covr^i \gets ECDF(\rho_{\dashv}^i, \mathbf{X_{e}})$
\State $Supp^i \gets ECDF([\rho_{\dashv}^i ; \rho_{\vdash}^i], [\mathbf{X_{e}}; \boldsymbol{\mu}])$ 
\State $Con^i \gets Supp^i/Covr^i$ 
\State $\alpha^i \gets weightedSum(Rel^i,Covr^i, Supp^i, Con^i)$ 
\EndFor
\State $\rho \gets FilterRules([\rho_{\dashv}^i ; \rho_{\vdash}^i], \alpha, t_\alpha)$ 
\State \textbf{return} $\rho$
\EndProcedure
\end{algorithmic}
\end{algorithm}

\subsubsection{Explanation Dataset Generation}
In the BO setting, a predefined dataset is not available because data is dynamically acquired during the optimization process. To overcome this, TNTRules generates an explanation dataset by uniformly sampling the search space $\mathcal{D}$, creating a set of $N_e$ samples: 
\[
\mathbf{X_{e}} = \{x_1, \ldots, x_{N_e}\} \sim \mathcal{U}(\mathcal{D}).
\]
The sampled points are queried with the GP model to obtain the posterior mean ($\boldsymbol{\mu}$) and standard deviation ($\boldsymbol{\sigma_y}$). The explanation dataset is then formulated as:
\[
\mathbf{E} = [\mathbf{X_{e}} ; \boldsymbol{\mu} ; \boldsymbol{\sigma_y}].
\]
This dataset serves as the foundation for clustering and rule extraction.

\subsubsection{Clustering and Variance-Based Pruning}
Hierarchical Agglomerative Clustering (HAC) is applied to the explanation dataset $\mathbf{E}$ using the Ward criterion~\cite{murtagh2012algorithms}. The resulting linkage matrix $\mathbf{E_{link}}$ encodes the hierarchical structure of the clusters. TNTRules employs a variance-based pruning mechanism for clustering. Clusters are merged only if the variance of their target values ($\text{Var}(\boldsymbol{\mu})$) is below a predefined threshold $t_s$. This approach ensures that clusters are localized and meaningful in the context of BO.

\subsubsection{Rule Construction}
For each cluster identified through variance pruning, TNTRules construct rules that describe the parameter ranges and their corresponding outcomes. The parameter bounds within the cluster define the antecedents of the rules:
\[
\rho_{\dashv} = [\text{min}(\mathbf{X^i_e}), \text{max}(\mathbf{X^i_e})].
\]
The consequents are defined using the GP posterior uncertainty:
\[
\rho_{\vdash} = [\boldsymbol{\mu^i} - 2 \boldsymbol{\sigma^i_y}, \boldsymbol{\mu^i} + 2 \boldsymbol{\sigma^i_y}].
\]
Each rule is represented in an interpretable ``IF-THEN'' format, where the antecedents describe the input parameter ranges, and the consequents predict the associated outcomes.

\subsubsection{Rule Ranking and Filtering}
To refine the rule set, TNTRules evaluates the quality of each rule based on four metrics:
\begin{enumerate}
    \item \textbf{Coverage ($Covr$):} The proportion of the search space covered by the rule's antecedent.
    \item \textbf{Support ($Supp$):} The alignment of the rule's antecedent with regions containing sampled data.
    \item \textbf{Confidence ($Con$):} The ratio of Support to Coverage, indicating the rule's reliability.
    \item \textbf{Relevance ($Rel$):} The maximum log-likelihood derived from the GP for data points matching the rule.
\end{enumerate}
An overall interestingness score $\alpha$ is calculated as a weighted sum of these metrics:
\[
\alpha = w_1 \cdot Covr + w_2 \cdot Supp + w_3 \cdot Con + w_4 \cdot Rel.
\]
Rules with $\alpha$ above a threshold $t_\alpha$ are retained for the final explanation set.

\subsubsection{Visualization of Explanations}
TNTRules presents the final rules in three distinct visualization modes to enhance interpretability:
\begin{enumerate}
    \item \textbf{Global Rules:} The rules are displayed in an ordered ``IF-THEN'' format, ranked by their interestingness scores. These rules provide a comprehensive overview of the parameter relationships across the entire search space.
    \item \textbf{Local Sensitivity Graphs:} For specific clusters, sensitivity graphs illustrate how variations in individual parameters affect the objective function. These graphs highlight which parameters are most influential and provide actionable insights for tuning.
    \item \textbf{Actionable Rules:} Tailored to specific use case of parameter tuning, these explanations recommend which parameters to adjust and their optimal ranges. For example, users are guided to focus on parameters requiring tuning while ignoring those already in their optimal states.
\end{enumerate}
The visualization ensures that both global trends and localized details are accessible, enabling users to understand the optimization process and make informed decisions. The actionable explanations simplify interpretation by abstracting non-essential details, especially in cases with complex parameter interactions.

\section{Natural Language} \label{appendix:sec:textualexp}
We used GPT-4 to generate our textual explanations. The prompt we used is given below:

\textbf{Input Prompt Example}
\textit{Based on the provided tuning recommendations, generate a textual explanation in natural language:}

\begin{enumerate}
    \item Clearly identify the parameters that should not be tuned (stated as \textit{``maintain stability''}).
    \item Highlight the parameters that should be fine-tuned, specifying the recommended ranges.
    \item Use clear and concise language for readability.
\end{enumerate}

\begin{itemize}
    \item \textbf{Tune:}
    \begin{itemize}
        \item Mass $M$: [70, 74]
        \item yolk-to-white ratio $ywr$: [0.6, 0.9]
    \end{itemize}
    \item \textbf{Do not tune:}
    \begin{itemize}
        \item $\lambda$, altitude $A$, egg temperature $T_{egg}$, yolk temperature $T_{yolk}$
    \end{itemize}
\end{itemize}

\textbf{Output Example}

\textit{Maintain stability for altitude ($A$), egg temperature ($Tegg$), lambda ($\lambda$), and yolk temperature ($T_{yolk}$).
Fine-tune Mass ($M$) between 70 and 74, and yolk-to-white ratio ($ywr$) between 0.6 and 0.9 for optimal performance.}

\section{Scenarios}
\label{appendix:sec:scenarios}

\subsection{Parameter Sensitivity Analysis}
\label{appendix:parameter-sensitivity}

\begin{figure}
    \centering
\includegraphics[width=0.6\textwidth,trim={1cm 1cm 1cm 1,8cm},clip]{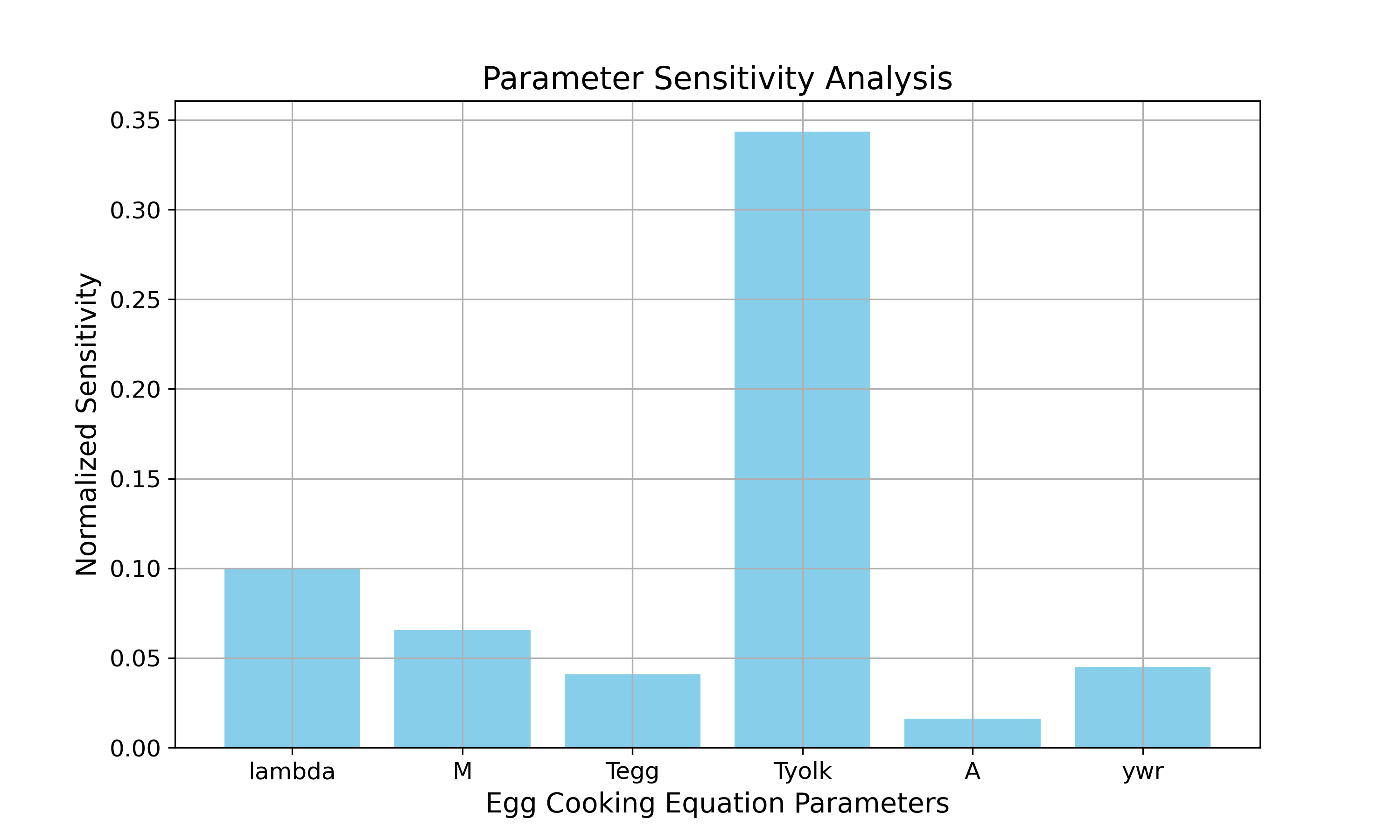}
    \caption{Our sensitivity analysis of the egg cooking formula for 10\% perturbation to the parameters identified Lambda and Tyolk as highly sensitive parameters. This outcome aligns with expectations, as these parameters are presumed stable in the original formulation, and any alterations significantly impact cooking time. To manage this sensitivity, we kept Lambda and Tyolk largely fixed in the tasks, ensuring that users do not need to deal with highly unstable settings. This approach helps maintain consistency and predictability in the study.}
    \label{fig:sensitivity-analysis}
\end{figure}

We performed a sensitivity analysis of the egg cooking formula to identify parameters that significantly influence cooking outcomes. This analysis allowed us to prioritize which parameters require precise control to ensure consistent and reproducible results (see Fig.~\ref{fig:sensitivity-analysis}). By understanding the sensitivity of these parameters, we could design cooking scenarios that are both realistic and adaptable, while minimizing the risk of task failure due to parameter variability. Specifically, we identified $\lambda$ and $T_{yolk}$ as critical parameters in the formulation. $\lambda$, the specific heat capacity, influences the heat transfer efficiency and varies depending on egg size and shell thickness. $T_{yolk}$, the temperature at the white-yolk boundary, determines the texture of the egg and is crucial for achieving desired results like soft boiled eggs. Out of the seven tasks, users were required to tune $\lambda$ and $T_{yolk}$ in only two cases. This design choice allowed us to incorporate scenarios involving sensitive parameters without making them the focus of most tasks. By doing so, we ensured diversity in the task set, allowing users to experience sensitive and less sensitive parameter tuning. This balance reflects realistic conditions where not all parameters are equally impactful while providing users with opportunities to engage with the most critical variables in some scenarios. 

\subsection{Feedback on Trials}
\label{appendix:egg-feedback}

The feedback was based on predefined time ranges: eggs cooked for \(4 \, \text{min} \, 20 \, \text{sec}\) to \(4 \, \text{min} \, 45 \, \text{sec}\) were considered ``perfect.'' Eggs ``slightly overcooked'' (\(4 \, \text{min} \, 45 \, \text{sec}\) to \(5 \, \text{min} \, 30 \, \text{sec}\)) or ``slightly undercooked'' (\(3 \, \text{min} \, 35 \, \text{sec}\) to \(4 \, \text{min} \, 20 \, \text{sec}\)) received corresponding feedback. Eggs cooked for less than \(3 \, \text{min} \, 35 \, \text{sec}\) were labeled ``undercooked,'' while those exceeding \(5 \, \text{min} \, 30 \, \text{sec}\) were labeled ``overcooked.'' These feedback ranges ensured that participants could receive meaningful guidance regardless of whether they achieved the perfect egg on their first attempt or required further tuning.

Table~\ref{tab:egg-feedback} shows the visual feedback.

\begin{table}[tb]
\centering
\caption{Feedback for participants, depending on how close their selected parameters got to the perfect parameter settings.  }
\begin{tabular}{lr cccc}  
\toprule
& \includegraphics[width=2cm]{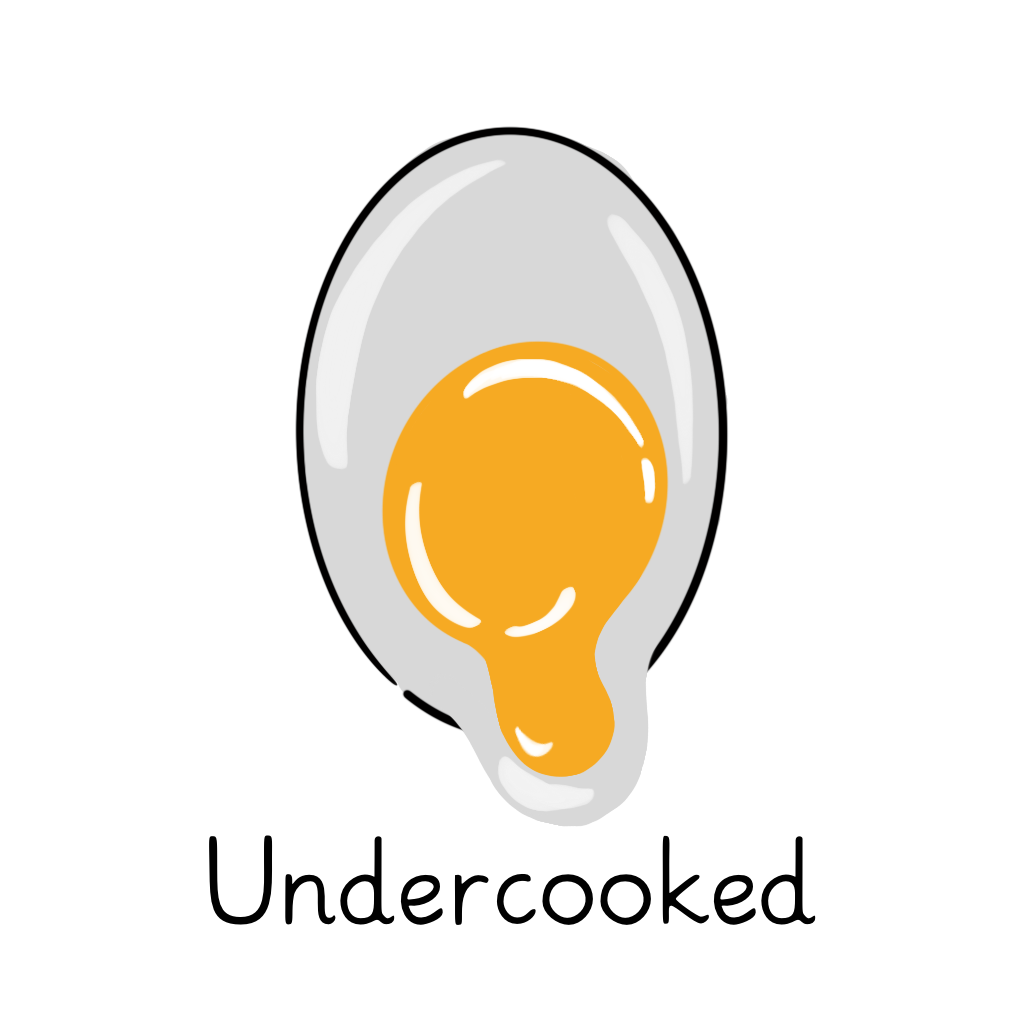}
& \includegraphics[width=2cm]{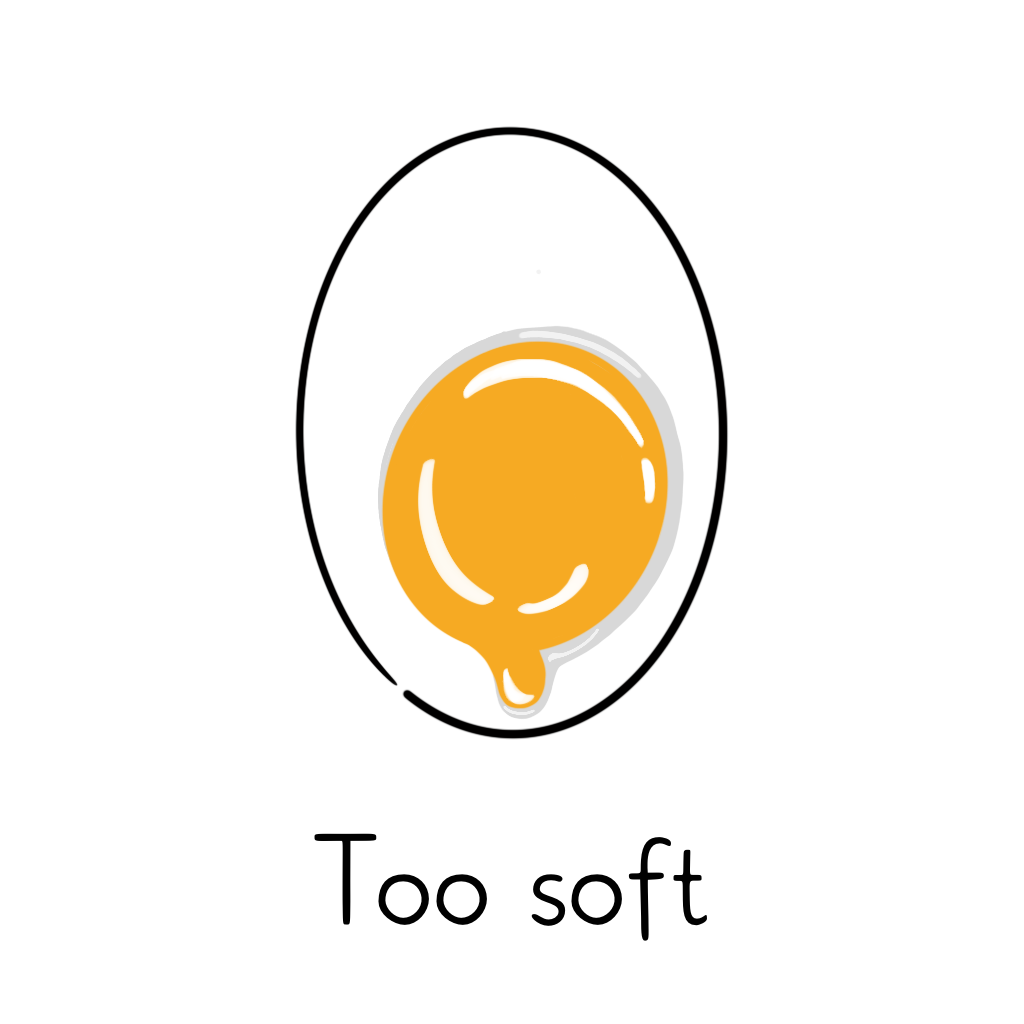}
& \includegraphics[width=2cm]{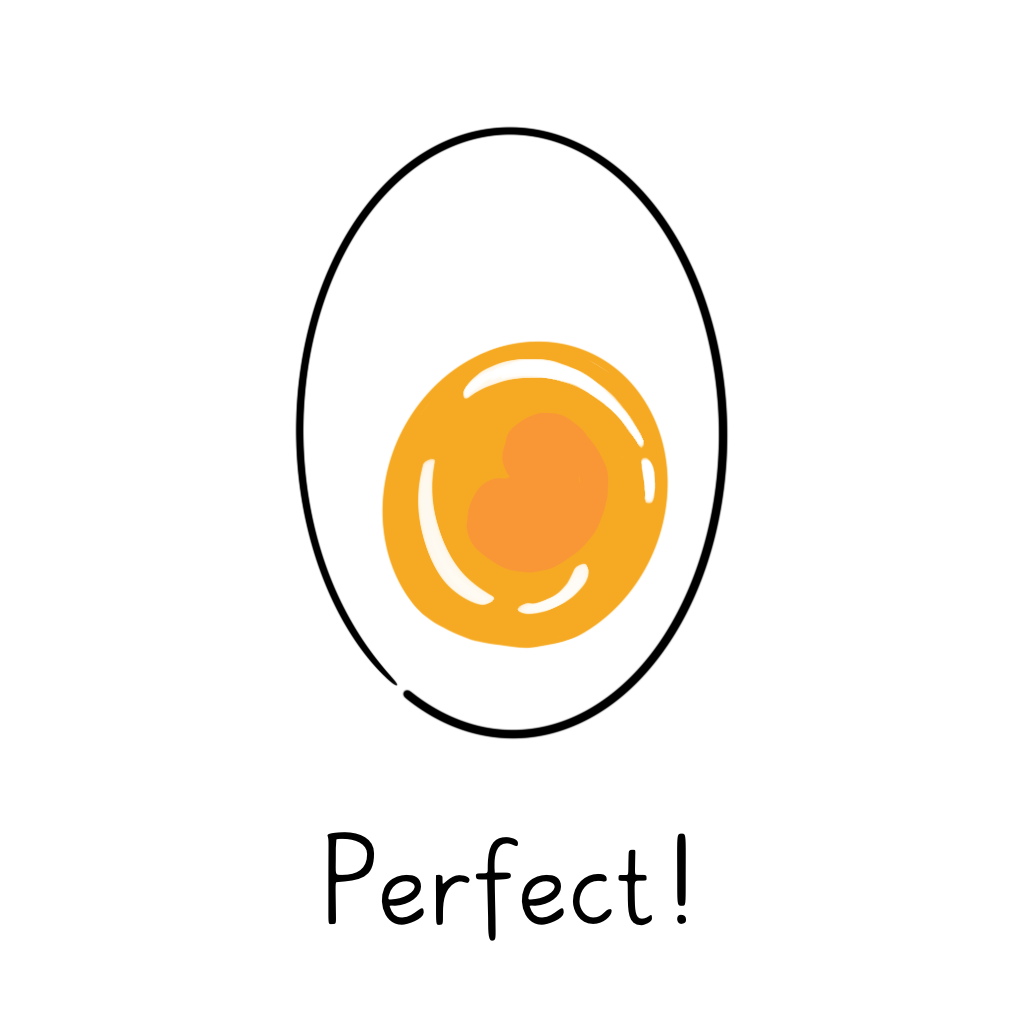}
& \includegraphics[width=2cm]{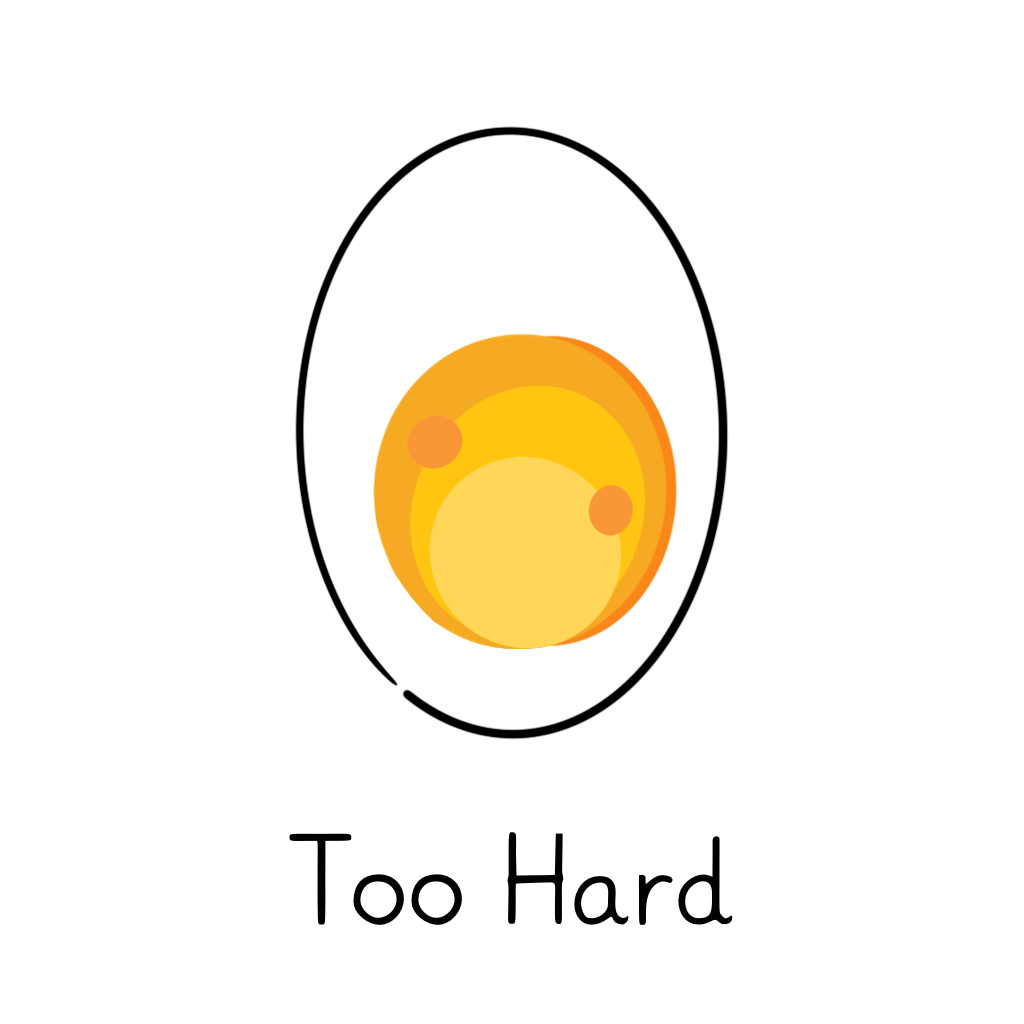}
& \includegraphics[width=2cm]{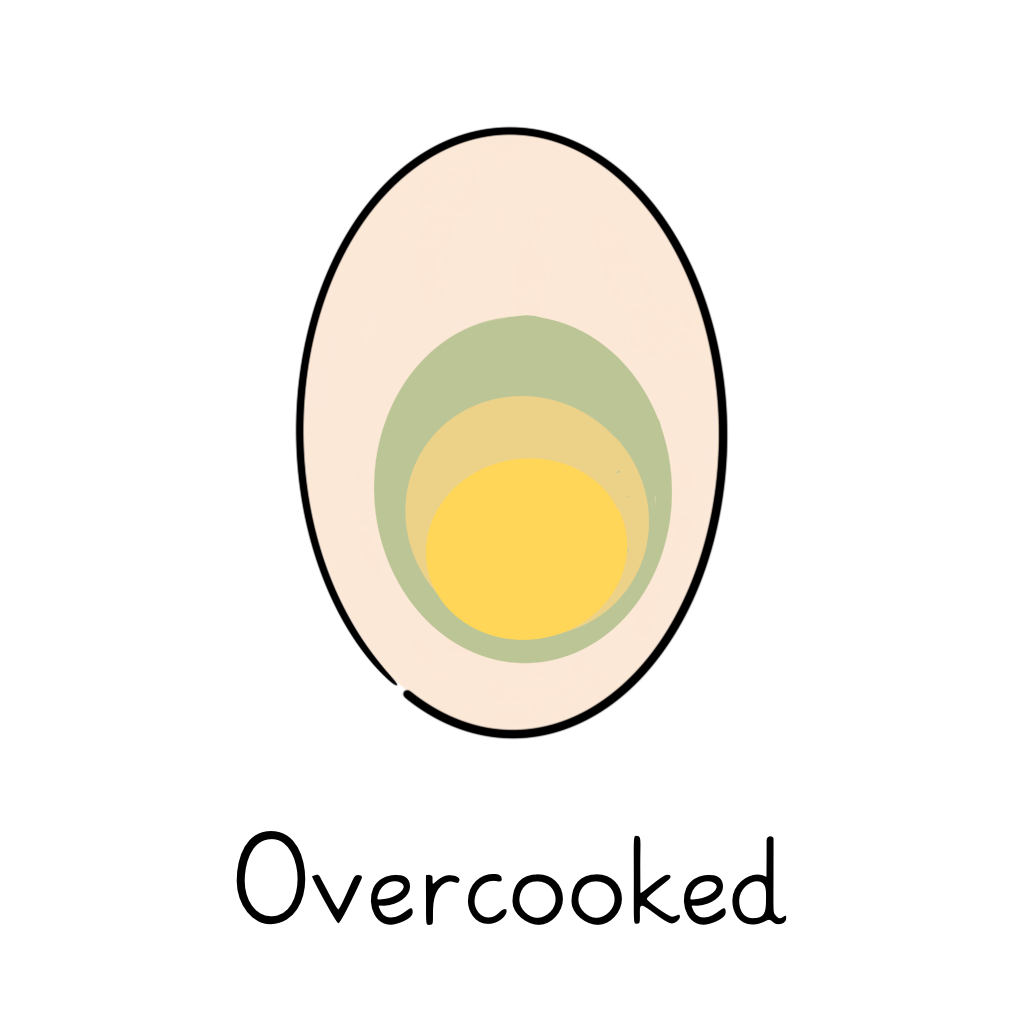}\\
\bottomrule
\end{tabular}

\label{tab:egg-feedback}
\end{table}

\section{Results}
\subsection{Participant Demographics} 
\label{appendix:demographics}
More details on participants: 

A majority of 135 participants~(63\%) had a university, followed by 50 participants~(23\%) with a high school degree and 16~(8\%) with a technical school diploma. Only six participants had a doctorate. The remaining four had completed middle-school and two participants had obtained no formal school leaving certificate. Notably, approximately half of the participants indicated that they used AI \textit{very frequently}~(23\%) or \textit{frequently}~(26\%). Sixty-one~(29\%) were occasional AI users. Only 16 participants each self-described their use of AI as \textit{rarely}, \textit{very rarely}, or \textit{never}~(8\% each). In contrast, only 8\% reported to \textit{very frequently} use explainable AI (XAI). Forty-one~(19\%) and fifty-nine~(28\%) participants were \textit{frequent} respectively \textit{occasional} XAI users. The other half \textit{rarely}~(13\%), \textit{very rarely}~(11\%), or \textit{never}~(21\%) used XAI. 
\subsection{Egg difficulty} \label{appendix:egg-difficulty}
A total of 1278 eggs were cooked in the study, 213 per egg type.
On average the difficulty of cooking the eggs was perceived as slightly difficult (mean between 4.5 and 5.0) for all egg types. Participants used the full range of the scale for all egg types (Min = 1: very easy, Max =  7: very difficult). An overview of the rating distributions is shown in Figure~\ref{fig:results:egg-diffiulty}.
The difficulty ratings for all egg types were not normally distributed (Shapiro-Wilk, all p<0.001). A Kruskal-Wallis test revealed significant differences between the rating means (W= 38.3, p=0.0000009934).
Only the goose egg (M=5.3, SD=2.0) was more difficult than the trial task of cooking a chicken egg (M=4.5, SD=1.7; post-hoc Dunn's test with Bonferroni correction (p =0.000007)). 
p-value for Chicken vs. Goose: 
\begin{figure}
    \centering
    \includegraphics[width=0.5\linewidth]{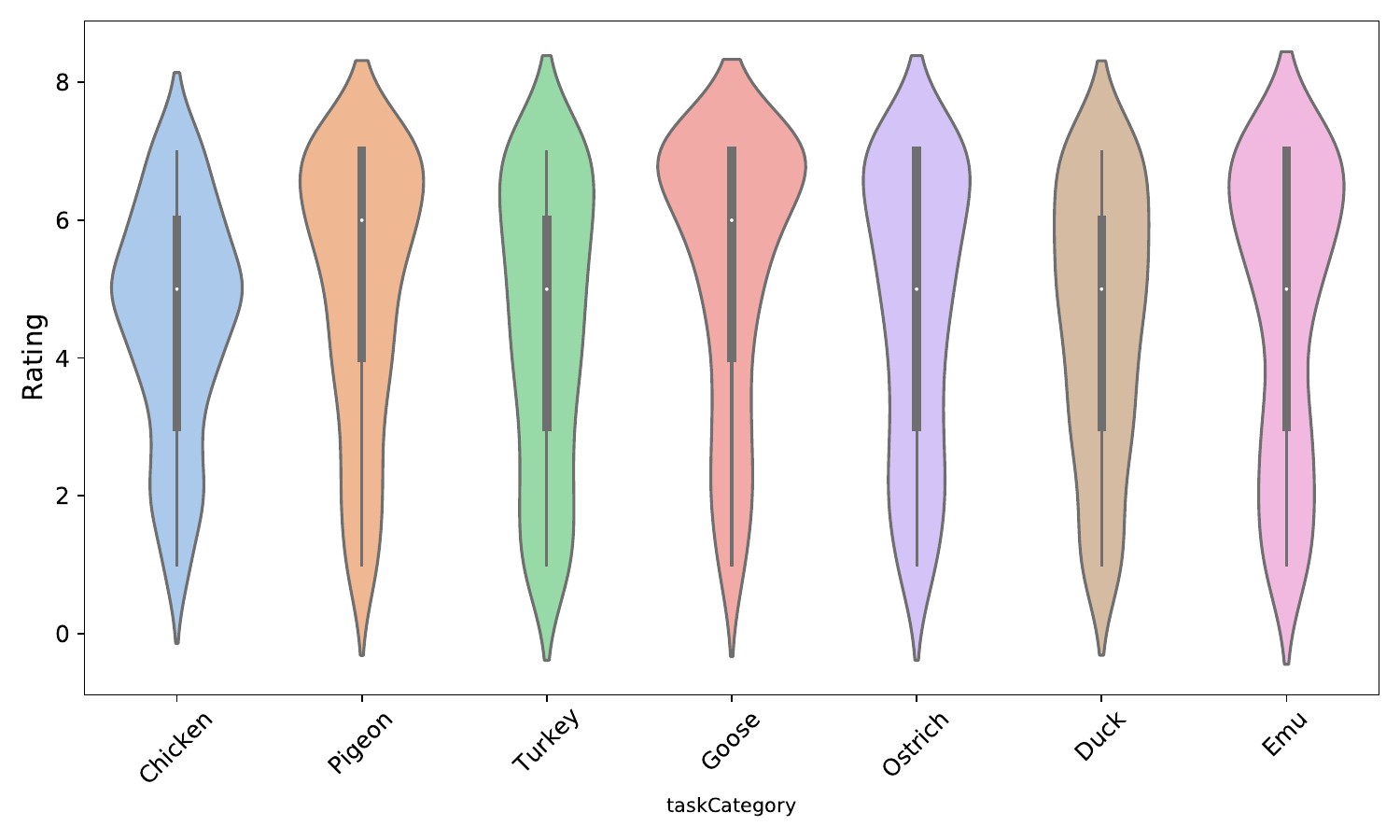}
    \caption{Egg difficulty, 1 - very easy, 7 - very difficult}
    \label{fig:results:egg-diffiulty}
\end{figure}

\subsection{Do explanations facilitate parameter tuning task?}
\begin{figure}[htbp]
    \centering
    \includegraphics[width=0.24\linewidth]{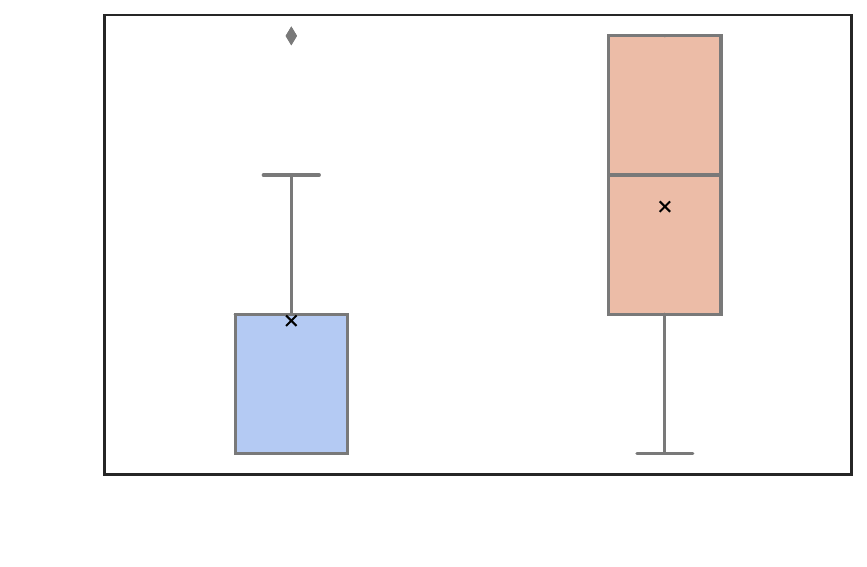}
    \includegraphics[width=0.24\linewidth]{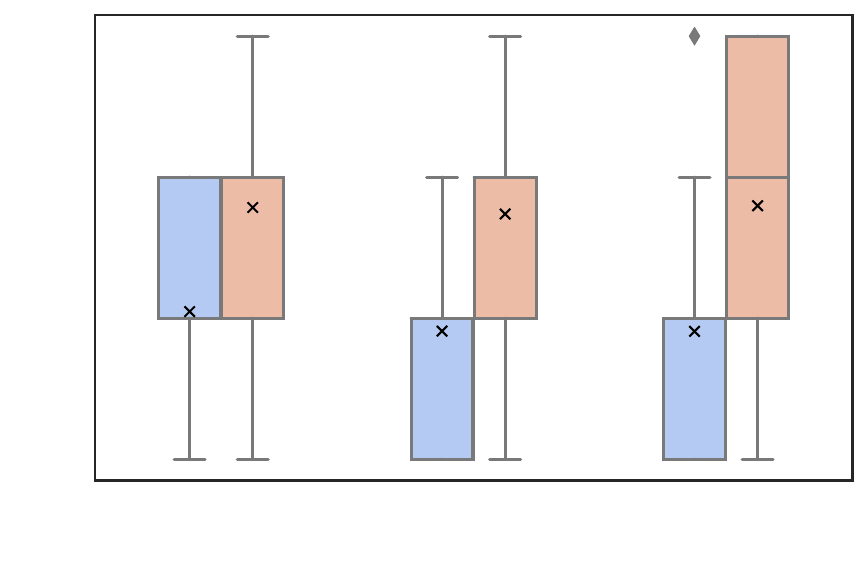}
    \includegraphics[width=0.24\linewidth]{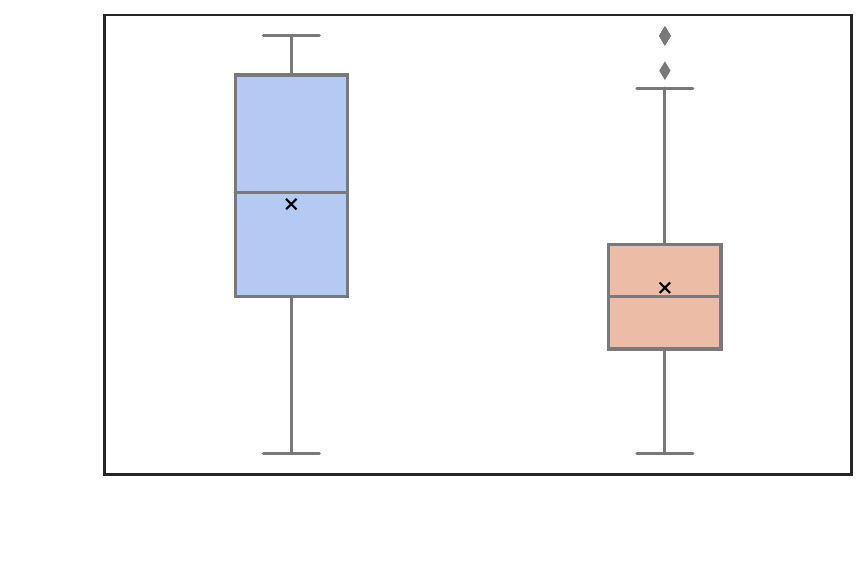}
    \includegraphics[width=0.24\linewidth]{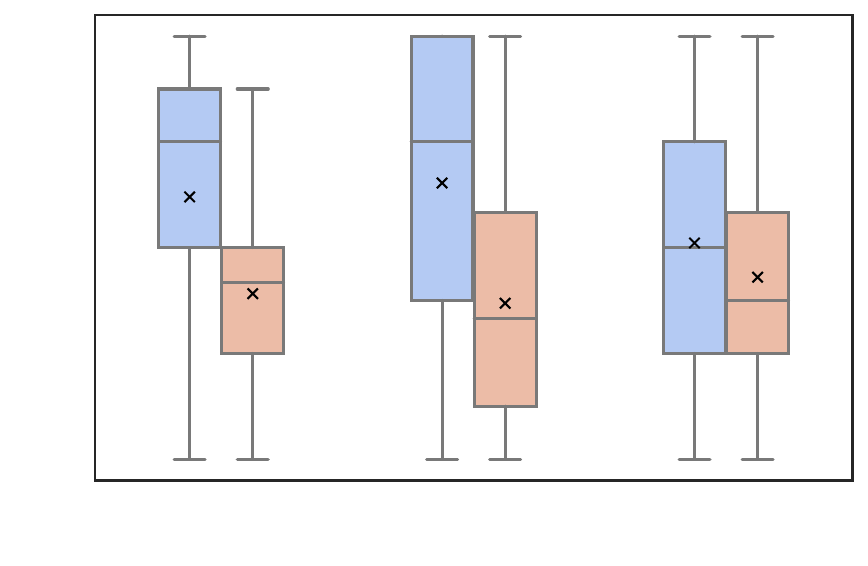}
    \caption{Performance, left success rate, right number of trials. \emph{x} on the plots represent the means. }
    \label{fig:results:performance}
\end{figure}

Fig.\ref{fig:results:performance} shows the distribution for task success and trial measures.

\subsection{Do explanations increase user’s trust in the system?} \label{appendix:trust}
\begin{figure}[htbp]
\centering
\subfigure[TIA sum score.]{
    \includegraphics[width=0.5\linewidth]{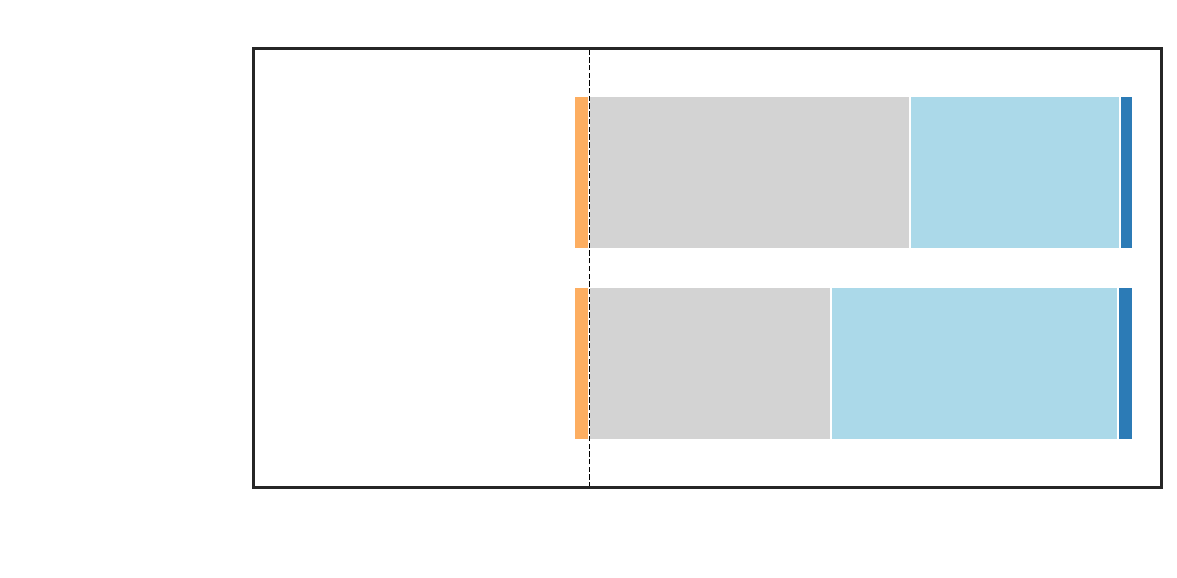}
    \includegraphics[width=0.5\linewidth]{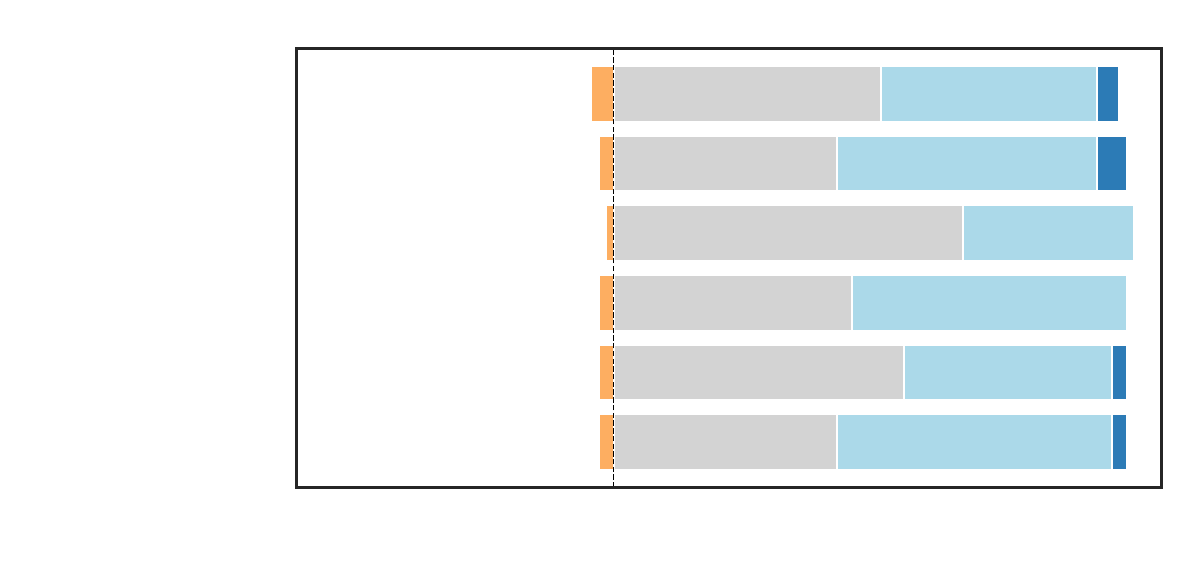}
}
\subfigure[TIA dimensions]{
    \includegraphics[width=0.5\linewidth]{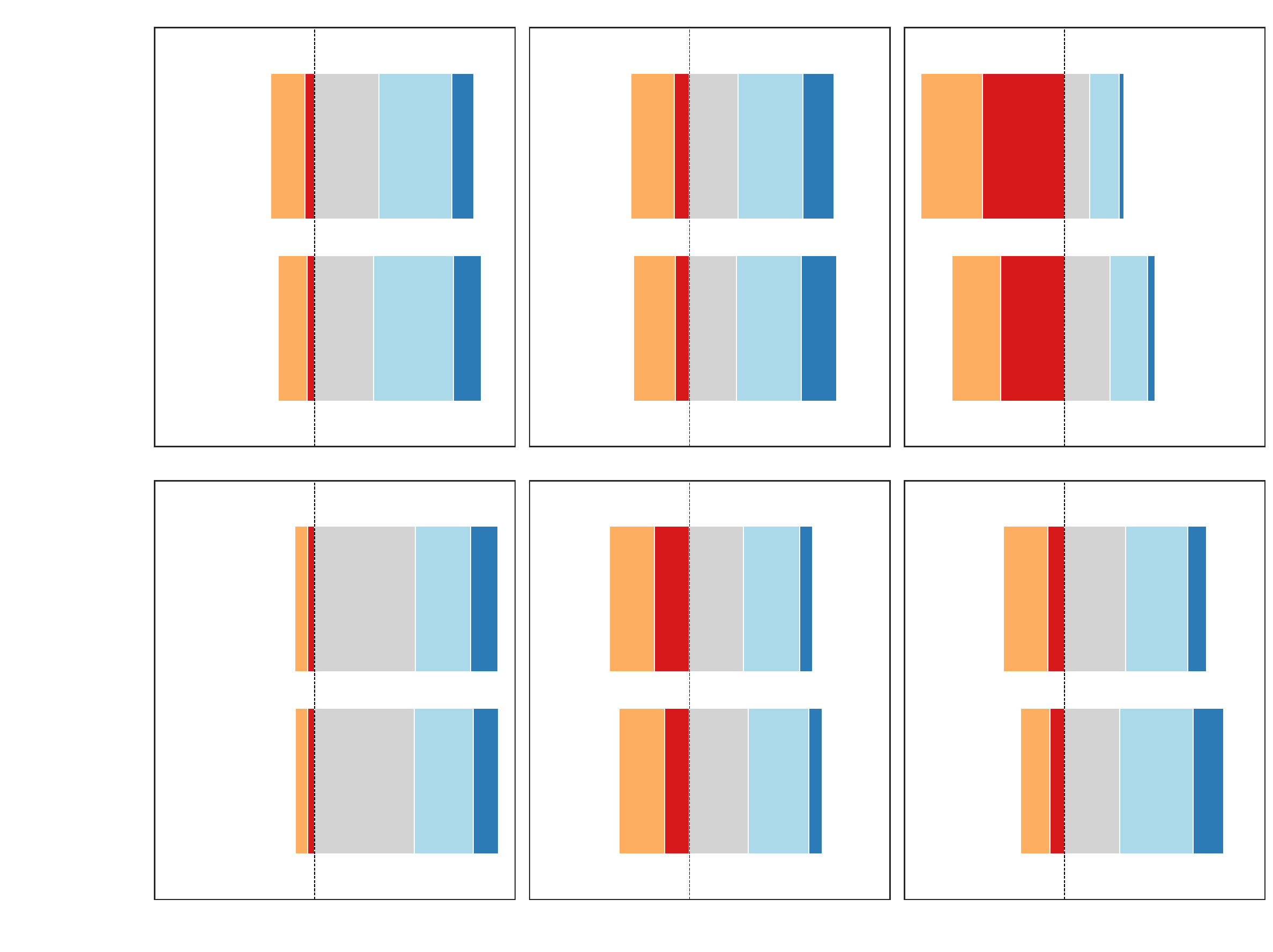}
    \includegraphics[width=0.5\linewidth]{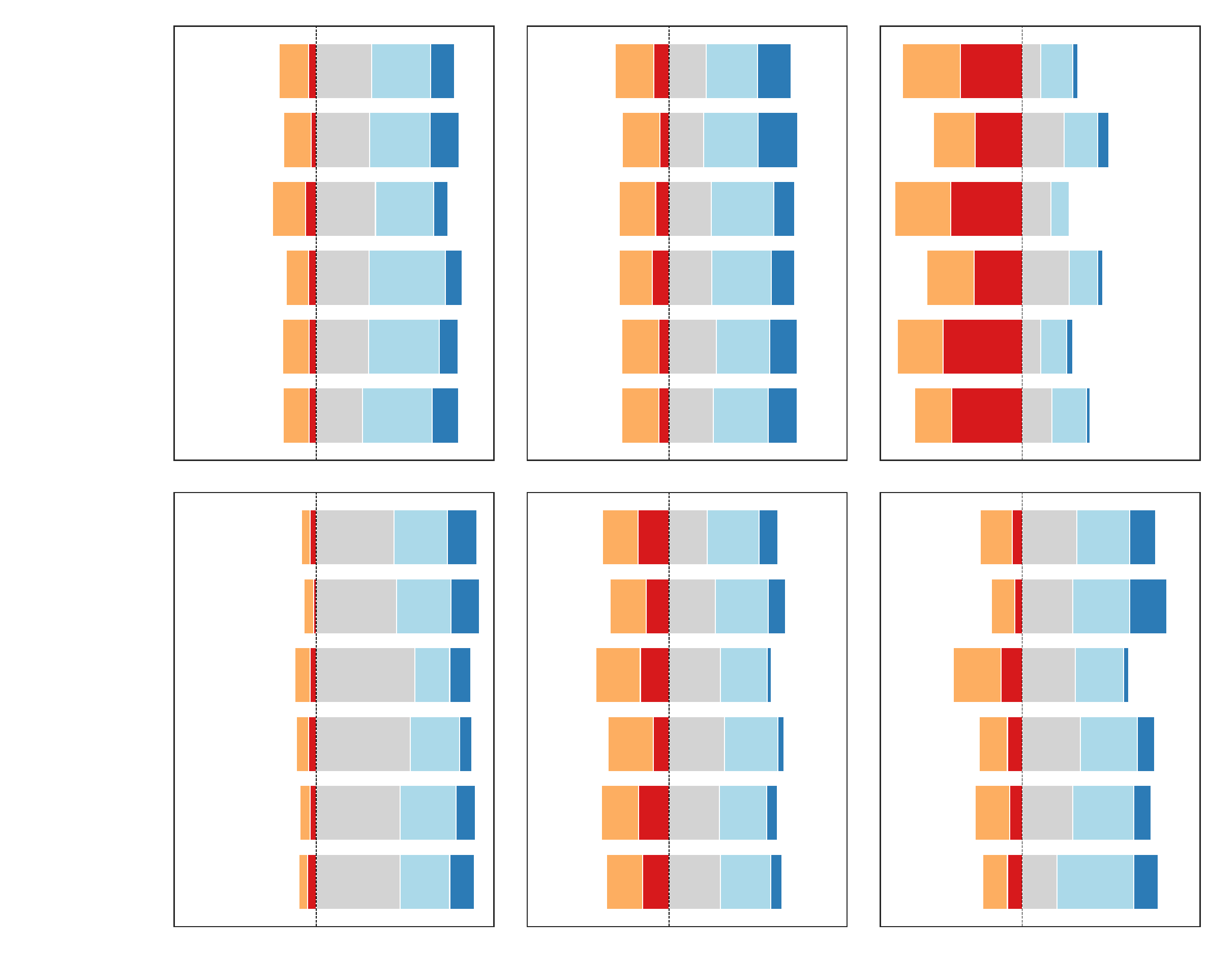}
}
\caption{Distribution of responses for the trust in automation (TIA) questionnaire. Colour legend: 
\protect\coloredsquare{cstronglyagree} strongly disagree, 
\protect\coloredsquare{coloragree} rather disagree, 
\protect\coloredsquare{colorneitheragree} neither agree nor disagree, 
\protect\coloredsquare{colordisagree} rather agree,
\protect\coloredsquare{colorstronglydisagree} strongly agree.
Left column: explanation vs. baseline (\xnone). Right: Comparing explanation formats (\xvis, \xrules, \xtext). There is no significant difference in mean values in all conditions.} 
\label{fig:results:trust}
\end{figure}
Detailed Likert scale results for the TiA questionnaire are presented in Fig.~\ref{fig:results:trust}. We observed that users significantly increased their trust in the system when explanations were provided.

\subsection{Do explanations increase user’s task load?} \label{appendix:task-load}
Detailed results for the NASA-TLX subscales and the raw TLX index are presented in Table~\ref{tab:nasa_tlx}. The findings demonstrate that various explanation formats offer distinct advantages, while consistently delivering significant benefits, such as reduced mental load, decreased effort, and improved performance. The availability of explanations positively impacted users without imposing additional cognitive load on them. Consequently, changes in the overall raw TLX index were statistically insignificant.

\begin{table}[ht]
  \caption{NASA-TLX subscales and RAW TLX index results comparing explanation types means and SDs. * ($p < .05$), ** ($p < .01$), *** ($p < .001$), and ns (non-significant) indicating significance levels. Significant reduction in mental demand and performance improvement for all explanation formats. \xrules show a reduction in frustration, and \xrules with \xvis show a reduction in effort as well.}
  \label{tab:nasa_tlx}
  \centering
  \scalebox{0.58}{
  \begin{tabular}{lllllllll}
    \toprule
    \textbf{Explanation} & \textbf{Phase} & \textbf{Mental Demand} & \textbf{Physical Demand} & \textbf{Temporal Demand} & \textbf{Performance} & \textbf{Effort} & \textbf{Frustration} & \textbf{RAW TLX Index} \\ 
    \midrule
    \xrules & Baseline & M=64.1, SD=24.5 * & M=21.6, SD=26.2 ns & M=27.9, SD=20.7 ns & M=33.8, SD=24.4 *** & M=60.8, SD=22.9 * & M=49.5, SD=31.6 ** & M=43.0, SD=13.8 ns \\
    \xrules & Treatment & M=59.4, SD=26.2 * & M=22.3, SD=28.7 ns & M=26.1, SD=22.1 ns & M=53.1, SD=26.2 *** & M=54.5, SD=25.5 * & M=40.4, SD=31.7 ** & M=42.6, SD=15.1 ns \\
    \xvis & Baseline & M=66.0, SD=27.7 * & M=21.6, SD=27.2 ns & M=32.9, SD=24.1 ns & M=37.8, SD=27.9 *** & M=66.7, SD=26.4 *** & M=47.7, SD=33.5 ns & M=45.4, SD=15.0 ns \\
    \xvis & Treatment & M=59.6, SD=30.3 * & M=23.9, SD=30.6 ns & M=32.4, SD=26.4 ns & M=56.6, SD=28.1 *** & M=57.0, SD=24.5 *** & M=39.9, SD=36.2 ns & M=44.9, SD=16.4 ns \\
    \xtext & Baseline & M=64.3, SD=21.7 ** & M=23.0, SD=27.8 ns & M=32.2, SD=26.6 ns & M=40.6, SD=27.0 ** & M=65.5, SD=19.4 ns & M=50.7, SD=29.6 ns & M=46.0, SD=14.0 ns \\
    \xtext & Treatment & M=56.8, SD=28.1 ** & M=21.4, SD=27.3 ns & M=30.1, SD=25.5 ns & M=53.5, SD=30.7 ** & M=61.5, SD=22.6 ns & M=44.6, SD=31.6 ns & M=44.6, SD=16.1 ns \\
    Combined & Baseline & M=64.8, SD=24.6 *** & M=22.1, SD=27.0 ns & M=31.0, SD=23.9 ns & M=37.4, SD=26.5 *** & M=64.3, SD=23.1 *** & M=49.3, SD=31.5 ns & M=44.8, SD=14.1 ns \\
    Combined & Treatment & M=58.6, SD=28.2 *** & M=22.5, SD=28.8 ns & M=29.5, SD=24.8 ns & M=54.4, SD=28.3 *** & M=57.7, SD=24.3 *** & M=41.6, SD=33.2 ns & M=44.1, SD=15.8 ns \\ 
    \bottomrule
  \end{tabular}}
\end{table}
\end{appendix}

\bibliographystyle{splncs04}
\bibliography{facct_reference}

\begin{thebibliography}{10}
\providecommand{\url}[1]{\texttt{#1}}
\providecommand{\urlprefix}{URL }
\providecommand{\doi}[1]{https://doi.org/#1}

\bibitem{adachi2024looping}
Adachi, M., Planden, B., Howey, D.A., Osborne, M.A., Orbell, S., Ares, N., Muandet, K., Chau, S.L.: Looping in the human collaborative and explainable bayesian optimization (2024)

\bibitem{adadi2018peeking}
Adadi, A., Berrada, M.: Peeking inside the black-box: a survey on explainable artificial intelligence (xai). IEEE access  \textbf{6},  52138--52160 (2018)

\bibitem{arora2022explain}
Arora, S., Pruthi, D., Sadeh, N., Cohen, W.W., Lipton, Z.C., Neubig, G.: Explain, edit, and understand: Rethinking user study design for evaluating model explanations. In: Proceedings of the AAAI Conference on Artificial Intelligence. vol.~36, pp. 5277--5285 (2022)

\bibitem{arrieta2020explainable}
Arrieta, A.B., D{\'\i}az-Rodr{\'\i}guez, N., Del~Ser, J., Bennetot, A., Tabik, S., Barbado, A., Garc{\'\i}a, S., Gil-L{\'o}pez, S., Molina, D., Benjamins, R., et~al.: Explainable artificial intelligence (xai): Concepts, taxonomies, opportunities and challenges toward responsible ai. Information fusion  \textbf{58},  82--115 (2020)

\bibitem{Breiman1984_classification}
Breiman, L., Friedman, J.H., Olshen, R.A., Stone, C.J.: Classification and Regression Trees. Wadsworth (1984)

\bibitem{10.1613/jair.1.12228}
Burkart, N., Huber, M.F.: A survey on the explainability of supervised machine learning. J. Artif. Int. Res.  \textbf{70},  245–317 (May 2021)

\bibitem{10.1145/3377325.3380623}
Burnett, M.: Explaining ai: fairly? well? In: Proceedings of the 25th International Conference on Intelligent User Interfaces. p. 1–2. IUI '20, Association for Computing Machinery, New York, NY, USA (2020)

\bibitem{casolin2024evaluatinginfluencesexplanationstyle}
Casolin, E., Salim, F.D., Newell, B.: Evaluating the influences of explanation style on human-ai reliance (2024), \url{https://arxiv.org/abs/2410.20067}

\bibitem{cezar2022domains}
Cezar-Vaz, M.R., Xavier, D.M., Bonow, C.A., Vaz, J.C., Cardoso, L.S., Sant’Anna, C.F., da~Costa, V.Z.: Domains of physical and mental workload in health work and unpaid domestic work by gender division: A study with primary health care workers in brazil. International journal of environmental research and public health  \textbf{19}, ~9816 (2022)

\bibitem{chakraborty2024explainable}
Chakraborty, T., Seifert, C., Wirth, C.: Explainable bayesian optimization (2024), \url{https://arxiv.org/abs/2401.13334}

\bibitem{chakraborty_post-hoc_nodate}
Chakraborty, T., Wirth, C., Seifert, C.: Post-hoc rule based explanations for black box bayesian optimization. In: Artificial Intelligence. ECAI 2023 International Workshops. pp. 320--337. Springer Nature Switzerland, Cham (2024)

\bibitem{chau2024explaining}
Chau, S.L., Muandet, K., Sejdinovic, D.: Explaining the uncertain: Stochastic shapley values for gaussian process models. Advances in Neural Information Processing Systems  \textbf{36} (2024)

\bibitem{Chen2019_this_looks}
Chen, C., Li, O., Tao, D., Barnett, A., Rudin, C., Su, J.K.: This looks like that: Deep learning for interpretable image recognition. In: Wallach, H., Larochelle, H., Beygelzimer, A., d\textquotesingle Alch\'{e}-Buc, F., Fox, E., Garnett, R. (eds.) Advances in Neural Information Processing Systems. vol.~32. Curran Associates, Inc. (2019), \url{https://proceedings.neurips.cc/paper_files/paper/2019/file/adf7ee2dcf142b0e11888e72b43fcb75-Paper.pdf}

\bibitem{Chen2016_infogan}
Chen, X., Duan, Y., Houthooft, R., Schulman, J., Sutskever, I., Abbeel, P.: Infogan: Interpretable representation learning by information maximizing generative adversarial nets. In: Lee, D., Sugiyama, M., Luxburg, U., Guyon, I., Garnett, R. (eds.) Advances in Neural Information Processing Systems. vol.~29. Curran Associates, Inc. (2016), \url{https://proceedings.neurips.cc/paper_files/paper/2016/file/7c9d0b1f96aebd7b5eca8c3edaa19ebb-Paper.pdf}

\bibitem{dinu2020challenging}
Dinu, J., Bigham, J., Zico~Kolter, J.: Challenging common interpretability assumptions in feature attribution explanations. In: NeurIPS Workshop: ML Retrospectives, Surveys \& Meta-Analyses (ML-RSA) (2020)

\bibitem{doshi2018considerations}
Doshi-Velez, F., Kim, B.: Considerations for evaluation and generalization in interpretable machine learning. Explainable and interpretable models in computer vision and machine learning pp. 3--17 (2018)

\bibitem{douglas2023dataquality}
Douglas, B.D., Ewell, P.J., Brauer, M.: Data quality in online human-subjects research: Comparisons between mturk, prolific, cloudresearch, qualtrics, and sona. PLOS ONE  \textbf{18}(3),  1--17 (03 2023)

\bibitem{faul2007g}
Faul, F., Erdfelder, E., Lang, A.G., Buchner, A.: G* power 3: A flexible statistical power analysis program for the social, behavioral, and biomedical sciences. Behavior research methods  \textbf{39}(2),  175--191 (2007)

\bibitem{feldhus2023-saliency}
Feldhus, N., Hennig, L., Nasert, M.D., Ebert, C., Schwarzenberg, R., M{\"o}ller, S.: Saliency map verbalization: Comparing feature importance representations from model-free and instruction-based methods. In: Dalvi~Mishra, B., Durrett, G., Jansen, P., Neves~Ribeiro, D., Wei, J. (eds.) Proceedings of the 1st Workshop on Natural Language Reasoning and Structured Explanations (NLRSE). pp. 30--46. Association for Computational Linguistics, Toronto, Canada (Jun 2023)

\bibitem{feurer2019hyperparameter}
Feurer, M., Hutter, F.: Hyperparameter optimization. Automated machine learning: Methods, systems, challenges pp. 3--33 (2019)

\bibitem{garnett_bayesoptbook_2023}
Garnett, R.: {Bayesian Optimization}. Cambridge University Press (2023)

\bibitem{grushetskaya2024hpexplorer}
Grushetskaya, Y., Sips, M., Schachtschneider, R., Saberioon, M., Mahan, A.: Hpexplorer: Xai method to explore the relationship between hyperparameters and model performance. In: Joint European Conference on Machine Learning and Knowledge Discovery in Databases. pp. 319--334. Springer (2024)

\bibitem{guidotti2018survey}
Guidotti, R., Monreale, A., Ruggieri, S., Turini, F., Giannotti, F., Pedreschi, D.: A survey of methods for explaining black box models. ACM computing surveys (CSUR)  \textbf{51}(5),  1--42 (2018)

\bibitem{doi:10.1126/scirobotics.aay7120}
Gunning, D., Stefik, M., Choi, J., Miller, T., Stumpf, S., Yang, G.Z.: Xai—explainable artificial intelligence. Science Robotics  \textbf{4}(37),  eaay7120 (2019)

\bibitem{Han2020_explaining}
Han, X., Wallace, B.C., Tsvetkov, Y.: Explaining black box predictions and unveiling data artifacts through influence functions. In: Jurafsky, D., Chai, J., Schluter, N., Tetreault, J. (eds.) Proceedings of the 58th Annual Meeting of the Association for Computational Linguistics. pp. 5553--5563. Association for Computational Linguistics, Online (Jul 2020)

\bibitem{hart2006nasa}
Hart, S.G.: Nasa-task load index (nasa-tlx); 20 years later. In: Proceedings of the human factors and ergonomics society annual meeting. vol.~50, pp. 904--908. Sage publications Sage CA: Los Angeles, CA (2006)

\bibitem{Hancock1988_nasa-tlx}
Hart, S.G., Staveland, L.E.: Development of nasa-tlx (task load index): Results of empirical and theoretical research. In: Hancock, P.A., Meshkati, N. (eds.) Human Mental Workload, Advances in Psychology, vol.~52, pp. 139--183. North-Holland (1988)

\bibitem{8933695}
Hohman, F., Srinivasan, A., Drucker, S.M.: Telegam: Combining visualization and verbalization for interpretable machine learning. In: 2019 IEEE Visualization Conference (VIS). pp. 151--155 (2019)

\bibitem{hvarfner2022pibo}
Hvarfner, C., Stoll, D., Souza, A., Nardi, L., Lindauer, M., Hutter, F.: \${\textbackslash}pi\${BO}: Augmenting acquisition functions with user beliefs for bayesian optimization. In: International Conference on Learning Representations (2022), \url{https://openreview.net/forum?id=MMAeCXIa89}

\bibitem{jirapinyo2017preclinical}
Jirapinyo, P., Abidi, W.M., Aihara, H., Zaki, T., Tsay, C., Imaeda, A.B., Thompson, C.C.: Preclinical endoscopic training using a part-task simulator: learning curve assessment and determination of threshold score for advancement to clinical endoscopy. Surgical endoscopy  \textbf{31},  4010--4015 (2017)

\bibitem{Kaptein2012_statistical-analysis-methody-for-CHI}
Kaptein, M., Robertson, J.: Rethinking statistical analysis methods for chi. In: Proceedings of the SIGCHI Conference on Human Factors in Computing Systems. p. 1105–1114. CHI '12, Association for Computing Machinery, New York, NY, USA (2012)

\bibitem{kim2024human}
Kim, J., Maathuis, H., Sent, D.: Human-centered evaluation of explainable ai applications: a systematic review. Frontiers in Artificial Intelligence  \textbf{7},  1456486 (2024)

\bibitem{korber2019theoretical}
K{\"o}rber, M.: Theoretical considerations and development of a questionnaire to measure trust in automation. In: Proceedings of the 20th Congress of the International Ergonomics Association (IEA 2018) Volume VI: Transport Ergonomics and Human Factors (TEHF), Aerospace Human Factors and Ergonomics 20. pp. 13--30. Springer (2019)

\bibitem{lersch2017towards}
Lersch, M.: Towards the perfect soft boiled egg (2017)

\bibitem{li2019explainability}
Li, M.Y., Adams, R.P.: Explainability constraints for bayesian optimization. In: 6th ICML Workshop on Automated Machine Learning (2019)

\bibitem{liao2022humancenteredexplainableaixai}
Liao, Q.V., Varshney, K.R.: Human-centered explainable ai (xai): From algorithms to user experiences (2022), \url{https://arxiv.org/abs/2110.10790}

\bibitem{lundberg2019explainable}
Lundberg, S.M., Erion, G.G., Chen, H., DeGrave, A.J., Prutkin, J.M., Nair, B., Katz, R., Himmelfarb, J., Bansal, N., Lee, S.: From local explanations to global understanding with explainable {AI} for trees. Nat. Mach. Intell.  \textbf{2}(1),  56--67 (2020)

\bibitem{MILLER20191}
Miller, T.: Explanation in artificial intelligence: Insights from the social sciences. Artificial Intelligence  \textbf{267},  1--38 (2019)

\bibitem{Miller_2021}
Miller, T.: Contrastive explanation: a structural-model approach. The Knowledge Engineering Review  \textbf{36}, ~e14 (2021)

\bibitem{moosbauer2024}
Moosbauer, J., Herbinger, J., Casalicchio, G., Lindauer, M., Bischl, B.: Explaining hyperparameter optimization via partial dependence plots. In: Proceedings of the 35th International Conference on Neural Information Processing Systems. NIPS '21, Curran Associates Inc., Red Hook, NY, USA (2021)

\bibitem{murtagh2012algorithms}
Murtagh, F., Contreras, P.: Algorithms for hierarchical clustering: an overview. Wiley Interdisciplinary Reviews: Data Mining and Knowledge Discovery  \textbf{2}(1),  86--97 (2012)

\bibitem{Nalepa2018}
Nalepa, G.J.: Rules as a Knowledge Representation Paradigm, pp. 3--25. Springer International Publishing, Cham (2018)

\bibitem{Nauta2023_pipnet}
Nauta, M., Schl{\"o}tterer, J., Van~Keulen, M., Seifert, C.: Pip-net: Patch-based intuitive prototypes for interpretable image classification. In: Proceedings of the IEEE/CVF Conference on Computer Vision and Pattern Recognition. pp. 2744--2753 (2023)

\bibitem{10.1145/3583558}
Nauta, M., Trienes, J., Pathak, S., Nguyen, E., Peters, M., Schmitt, Y., Schl\"{o}tterer, J., van Keulen, M., Seifert, C.: From anecdotal evidence to quantitative evaluation methods: A systematic review on evaluating explainable ai. ACM Comput. Surv.  (2023)

\bibitem{neumann2019data}
Neumann-Brosig, M., Marco, A., Schwarzmann, D., Trimpe, S.: Data-efficient autotuning with bayesian optimization: An industrial control study. IEEE Transactions on Control Systems Technology  \textbf{28}(3),  730--740 (2019)

\bibitem{Nguyen2024_llms}
Nguyen, V.B., Youssef, P., Seifert, C., Schl{\"o}tterer, J.: {LLM}s for generating and evaluating counterfactuals: A comprehensive study. In: Al-Onaizan, Y., Bansal, M., Chen, Y.N. (eds.) Findings of the Association for Computational Linguistics: EMNLP 2024. pp. 14809--14824. Association for Computational Linguistics, Miami, Florida, USA (Nov 2024)

\bibitem{nimmo2024user}
Nimmo, R., Constantinides, M., Zhou, K., Quercia, D., Stumpf, S.: User characteristics in explainable ai: The rabbit hole of personalization? In: Proceedings of the 2024 CHI Conference on Human Factors in Computing Systems. pp. 1--13 (2024)

\bibitem{openai2024gpt4technicalreport}
OpenAI: Gpt-4 technical report (2024), \url{https://arxiv.org/abs/2303.08774}

\bibitem{papenmeier2022complicated}
Papenmeier, A., Kern, D., Englebienne, G., Seifert, C.: It’s complicated: The relationship between user trust, model accuracy and explanations in ai. ACM Trans. Comput.-Hum. Interact.  \textbf{29}(4) (Mar 2022)

\bibitem{doi:10.1080/17579961.2024.2313795}
Pavlidis, G.: Unlocking the black box: analysing the eu artificial intelligence act’s framework for explainability in ai. Law, Innovation and Technology  \textbf{16}(1),  293--308 (2024)

\bibitem{pedro2020development}
Pedro, A., Pham, H.C., Kim, J.U., Park, C.: Development and evaluation of context-based assessment system for visualization-enhanced construction safety education. International Journal of Occupational Safety and Ergonomics  \textbf{26},  811--823 (2020)

\bibitem{prolific}
{Prolific}: {Prolific}. \url{https://www.prolific.com} (2024), version used: Current month(s) and year(s)

\bibitem{Quinlan1986_induction}
Quinlan, J.R.: Induction of decision trees. Mach. Learn.  \textbf{1}(1),  81--106 (1986)

\bibitem{rago2024exploring}
Rago, A., Palfi, B., Sukpanichnant, P., Nabli, H., Vivek, K., Kostopoulou, O., Kinross, J., Toni, F.: Exploring the effect of explanation content and format on user comprehension and trust (2024), \url{https://arxiv.org/abs/2408.17401}

\bibitem{Rasmussen2005-yj}
Rasmussen, C.E., Williams, C.K.I.: Gaussian processes for machine learning. Adaptive computation and machine learning, {MIT} Press (2006)

\bibitem{ribeiro2016why}
Ribeiro, M.T., Singh, S., Guestrin, C.: "why should {I} trust you?": Explaining the predictions of any classifier. In: Proceedings of the International Conference on Knowledge Discovery and Data Mining, {SIGKDD}. pp. 1135--1144. {ACM} (2016)

\bibitem{ribeiro2018anchors}
Ribeiro, M.T., Singh, S., Guestrin, C.: Anchors: High-precision model-agnostic explanations. In: Proceedings of the AAAI conference on artificial intelligence. vol.~32 (2018)

\bibitem{Robertson2012_likert-type-scales}
Robertson, J.: Likert-type scales, statistical methods, and effect sizes. Commun. ACM  \textbf{55}(5),  6–7 (May 2012)

\bibitem{rodemann2022accounting}
Rodemann, J., Augustin, T.: Accounting for gaussian process imprecision in bayesian optimization. In: International Symposium on Integrated Uncertainty in Knowledge Modelling and Decision Making. pp. 92--104. Springer (2022)

\bibitem{rodemann2024explaining}
Rodemann, J., Croppi, F., Arens, P., Sale, Y., Herbinger, J., Bischl, B., Hüllermeier, E., Augustin, T., Walsh, C.J., Casalicchio, G.: Explaining bayesian optimization by shapley values facilitates human-ai collaboration (2024)

\bibitem{10316181}
Rong, Y., Leemann, T., Nguyen, T.T., Fiedler, L., Qian, P., Unhelkar, V., Seidel, T., Kasneci, G., Kasneci, E.: Towards human-centered explainable ai: A survey of user studies for model explanations. IEEE Transactions on Pattern Analysis and Machine Intelligence  \textbf{46},  2104--2122 (2024)

\bibitem{rudin2019stop}
Rudin, C.: Stop explaining black box machine learning models for high stakes decisions and use interpretable models instead. Nature machine intelligence  \textbf{1}(5),  206--215 (2019)

\bibitem{rudin2022black}
Rudin, C.: Why black box machine learning should be avoided for high-stakes decisions, in brief. Nature Reviews Methods Primers  \textbf{2}(1), ~81 (2022)

\bibitem{Sauro2009_Post-Task-Usability-Questionaire}
Sauro, J., Dumas, J.S.: Comparison of three one-question, post-task usability questionnaires. In: Proceedings of the SIGCHI Conference on Human Factors in Computing Systems. p. 1599–1608. CHI '09, Association for Computing Machinery, New York, NY, USA (2009)

\bibitem{schuff2022human}
Schuff, H., Jacovi, A., Adel, H., Goldberg, Y., Vu, N.T.: Human interpretation of saliency-based explanation over text. In: Proceedings of the 2022 ACM Conference on Fairness, Accountability, and Transparency. pp. 611--636 (2022)

\bibitem{senoner2024explainable}
Senoner, J., Schallmoser, S., Kratzwald, B., Feuerriegel, S., Netland, T.: Explainable ai improves task performance in human--ai collaboration. Scientific Reports  \textbf{14}(1),  31150 (2024)

\bibitem{shahriari2015taking}
Shahriari, B., Swersky, K., Wang, Z., Adams, R.P., De~Freitas, N.: Taking the human out of the loop: A review of bayesian optimization. Proceedings of the IEEE  \textbf{104}(1),  148--175 (2015)

\bibitem{SHI2024104332}
Shi, Y., Chen, S., Liu, Y., Liu, J., Xuan, L., Li, G., Li, J., Zheng, J.: Towards the perfect soft-boiled chicken eggs: Defining cooking conditions, quality criteria, and safety assessments. Poultry Science p. 104332 (2024)

\bibitem{10.1145/3397481.3450662}
Szymanski, M., Millecamp, M., Verbert, K.: Visual, textual or hybrid: the effect of user expertise on different explanations. In: Proceedings of the 26th International Conference on Intelligent User Interfaces. p. 109–119. IUI '21, Association for Computing Machinery, New York, NY, USA (2021)

\bibitem{Tintarev2011}
Tintarev, N., Masthoff, J.: Designing and Evaluating Explanations for Recommender Systems, pp. 479--510. Springer US, Boston, MA (2011)

\bibitem{vasconcelos2023explanationsreduceoverrelianceai}
Vasconcelos, H., Jörke, M., Grunde-McLaughlin, M., Gerstenberg, T., Bernstein, M., Krishna, R.: Explanations can reduce overreliance on ai systems during decision-making (2023)

\bibitem{van2021evaluating}
van~der Waa, J., Nieuwburg, E., Cremers, A.H.M., Neerincx, M.A.: Evaluating {XAI:} {A} comparison of rule-based and example-based explanations. Artif. Intell.  \textbf{291},  103404 (2021)

\bibitem{10.1145/3290605.3300831}
Wang, D., Yang, Q., Abdul, A., Lim, B.Y.: Designing theory-driven user-centric explainable ai. In: Proceedings of the 2019 CHI Conference on Human Factors in Computing Systems. p. 1–15. CHI '19, Association for Computing Machinery, New York, NY, USA (2019)

\bibitem{wang2015falling}
Wang, F., Rudin, C.: Falling rule lists. In: Artificial intelligence and statistics. pp. 1013--1022. PMLR (2015)

\bibitem{WANG2022100728}
Wang, K., Dowling, A.W.: Bayesian optimization for chemical products and functional materials. Current Opinion in Chemical Engineering  \textbf{36},  100728 (2022)

\bibitem{Wang2024GradientBF}
Wang, Y., Zhang, T., Guo, X., Shen, Z.: Gradient based feature attribution in explainable ai: A technical review. ArXiv  \textbf{abs/2403.10415} (2024), \url{https://api.semanticscholar.org/CorpusID:268510511}

\bibitem{wanous1997overall}
Wanous, J.P., Reichers, A.E., Hudy, M.J.: Overall job satisfaction: how good are single-item measures? Journal of applied Psychology  \textbf{82}(2), ~247 (1997)

\end{thebibliography}

\end{document}